\documentclass[superscriptaddress,floatfix]{iopart}

\usepackage{iopams}
\usepackage[T1]{fontenc}
\usepackage[utf8]{inputenc}
\usepackage{graphicx}% Include figure files
\usepackage{xcolor}
\usepackage{hyperref}% add hypertext capabilities
\usepackage{amssymb}
\usepackage{setstack}
\usepackage{siunitx}

\usepackage{siunitx}

\newcommand{\sub}[1]{_{\mathrm{#1}}}
\newcommand{\Sn}{\mathcal{S}}
\newcommand{\inv}{^{-1}}

\newcommand{\B}{\mathcal{B}}
\newcommand{\Lr}{\mathcal{L}}
\newcommand{\sky}{\vec{n}}

\newcommand{\weak}{\sub{weak}}

\newcommand{\sft}{\sub{sft}}
\newcommand{\seg}{\sub{seg}}
\newcommand{\Tspan}{T\sub{span}}
\newcommand{\Ndet}{N\sub{d}}
\newcommand{\dfact}{\gamma}

\newcommand{\expect}[1]{E\left[#1\right]}
\newcommand{\cov}[2]{\mathrm{cov}\left[#1, #2\right]}
\newcommand{\var}[1]{\mathrm{var}\left[#1\right]}
\newcommand{\mean}[1]{\overline{#1}}

\newcommand{\I}{\mathbb{I}}
\newcommand{\transpose}{^{\mathrm{T}}}
\newcommand{\diag}[1]{\mathrm{diag}\left(#1\right)}
\newcommand{\trace}{\mathrm{Tr}}
\newcommand{\M}{\mathcal{M}}
\newcommand{\R}{\mathcal{R}}
\newcommand{\W}{\mathcal{W}}
\newcommand{\Q}{\mathcal{Q}}
\newcommand{\plus}{{+}}
\newcommand{\cross}{{\times}}
\newcommand{\Aplus}{A_\plus}
\newcommand{\Across}{A_\cross}

\newcommand{\dop}{\lambda}
\newcommand{\pluscross}{{+,\times}}
\newcommand{\uvec}[1]{\boldsymbol{#1}}
\newcommand{\Amp}{\mathcal{A}}
\newcommand{\vAmp}{\uvec{\Amp}}

\newcommand{\DET}{\sub{det}}
\newcommand{\thresh}{\sub{thr}}
\newcommand{\F}{\mathcal{F}}

\newcommand{\Tdata}{T\sub{data}}
\newcommand{\Tsft}{T\sub{sft}}
\newcommand{\av}[1]{\langle{#1}\rangle}

\newcommand{\gchi}{\tilde{\chi}}
\newcommand{\hO}{h_0}
\newcommand{\cosi}{\eta}
\newcommand{\phiO}{\phi_0}
\newcommand{\sco}[1]{\widehat{#1}}

\newcommand{\FA}{\sub{fa}}
\newcommand{\prob}[2]{P\left(#1|#2\right)}
\newcommand{\prior}[1]{P\left(#1\right)}

\newcommand{\stat}{S}

\newcommand{\Gauss}{{\mathrm{G}}}

\newcommand{\Signal}{{\mathrm{S}}}

\newcommand{\Hyp}{\mathcal{H}}
\newcommand{\HypS}{\Hyp_\Signal}

\newcommand{\HypG}{\Hyp_\Gauss}

\newcommand{\Ord}[1]{\mathcal{O}\left(#1\right)}

\newcommand{\scalar}[2]{\left(#1|#2\right)}

\newcommand{\vx}{\uvec{x}}

\newcommand{\vz}{\uvec{z}}
\newcommand{\vs}{\uvec{s}}
\newcommand{\vn}{\uvec{n}}

\newcommand{\rel}{_{*}}
\newcommand{\hrel}{h\rel}
\newcommand{\Hrel}{H\rel}

\newcommand{\rhog}{g}

\newcommand{\bcoh}{\beta}
\newcommand{\bsco}{\sco{\beta}}
\newcommand{\AmuPriorConst}{k}
\newcommand{\hOPriorConst}{\kappa}
\newcommand{\BeroWhelan}{{\sub{BW}}}
\newcommand{\cpf}[1]{\check{#1}}
\newcommand{\mle}[1]{\hat{#1}}

\newcommand{\chisq}{{}\sub{\chi^2}}

\newcommand{\tens}[1]{\underline{#1}}
\newcommand{\five}[1]{\check{#1}}
\newcommand{\sid}{_{\oplus}}

\newcommand{\eqref}{\eref}
\newcommand{\commitDATE}{2026-02-26 11:07:13 +0100}

\newcommand{\commitIDshort}{commitID: 5f6433a}
\newcommand{\commitSTATUS}{CLEAN}

\begin{document}

\title{Analytic weak-signal approximation of the Bayes factor for continuous gravitational waves}

\author{Reinhard Prix$^{1,2}$}
\address{$^1$ Max-Planck-Institute for Gravitational Physics (Albert-Einstein-Institute), D-30167
  Hannover, Germany}
\address{$^2$ Leibniz University Hannover, D-30167 Hannover, Germany}
\ead{Reinhard.Prix@aei.mpg.de}

\begin{abstract}
  We generalize the targeted $\B$-statistic for continuous gravitational waves by modeling the
  $\hO$-prior as a half-Gaussian distribution with scale parameter $H$.
  This approach retains analytic tractability for two of the four amplitude marginalization
  integrals and recovers the standard $\B$-statistic in the strong-signal limit ($H\rightarrow\infty$).
  By Taylor-expanding the weak-signal regime ($H\rightarrow 0$), the new prior enables fully
  analytic amplitude marginalization, resulting in a simple, explicit statistic that is as
  computationally efficient as the maximum-likelihood $\F$-statistic, but significantly more robust.
  Numerical tests show that for day-long coherent searches, the weak-signal Bayes factor
  achieves sensitivities comparable to the $\F$-statistic, though marginally lower than the standard
  $\B$-statistic (and the Bero-Whelan approximation).
  In semi-coherent searches over short (compared to a day) segments, this approximation matches or
  outperforms the weighted dominant-response $\sco{\F}\sub{ABw}$-statistic and returns to the
  sensitivity of the (weighted) $\sco{\F}\sub{w}$-statistic for longer segments.
  Overall the new Bayes-factor approximation demonstrates state-of-the-art or improved sensitivity
  across a wide range of segment lengths we tested (from $\SI{900}{\second}$ to $\SI{10}{days}$).

  \vspace*{0.5cm}
  [{\commitDATE; \commitIDshort-\commitSTATUS}]
\end{abstract}

% FIXME: uncomment for submitted version
% \submitto{\CQG}

% \showthe\textwidth
% \showthe\columnwidth
% \makeatletter
% \show\f@size
% \makeatother
%\maketitle

\section{Introduction}
\label{sec:introduction}

Continuous gravitational waves (CWs), expected to be emitted by spinning non-axisymmetric neutron
stars in our galaxy, are one of the most anticipated but still undetected types of gravitational
waves.
Finding CWs will mark the culmination of decades of research aimed at refining search methods
and applying them to data from ground-based detectors, such as LIGO Hanford (H1), Livingston (L1)
\cite{collaboration_advanced_2015}, and Virgo (V1) \cite{acernese_advanced_2014}.
For recent reviews, see, for example, \cite{riles_searches_2023,wette_searches_2023}.

CW signals manifest in the detector data as amplitude-modulated quasi-sinusoidal waveforms with
slowly varying (generally decreasing) frequency.  These are long-lasting but extremely weak signals
compared to the detector noise, and will therefore require combining months to years of data to become
detectable.
The signals are characterized by four \emph{amplitude} parameters: the overall amplitude
$\hO$, two polarization angles $\iota$ and $\psi$, and the initial phase $\phiO$.
An additional set of \emph{phase-evolution} parameters, such as the frequency (and its time
derivatives), sky position and binary-orbital parameters, is required to fully determine the signal,
depending on the search space considered.

A key research focus for detecting CWs is the development of statistics that maximize the detection
probability at a fixed false-alarm level, while being as computationally efficient as possible.
A significant milestone in this area was the seminal work by Jaranowski, Królak, and Schutz
\cite{jks98:_data} (henceforth ``JKS''), demonstrating that the likelihood ratio can be analytically
maximized over the four unknown amplitude parameters after a coordinate transformation.
This results in the $\F$-statistic, which only requires explicit searches over phase-evolution
parameters, thereby substantially reducing the computing cost.
Another closely-related coherent statistic was constructed by different arguments within the 5-vector
formalism \cite{astone_method_2010}, equally dispensing with the amplitude parameters and resulting
in very similar detection power (see \ref{sec:relation-5-vector} for the detailed relation).

Searle pointed out in \cite{searle_monte-carlo_2008} that for composite signal hypotheses (i.e.,
those with unknown parameters), the Neyman-Pearson lemma shows that the optimal statistic is the
\emph{marginalized} (rather than maximized) likelihood ratio, also known as the Bayes factor.
This was illustrated in the CW context in \cite{2009CQGra..26t4013P}, showing that marginalizing
with more physical priors over the amplitude parameters resulted in a (slightly) more powerful
statistic (referred to as the $\B$-statistic) than the $\F$-statistic.
However, this statistic entails a substantially higher computing cost, given that only two of the
four marginalization integrals have been solved analytically \cite{2014CQGra..31f5002W}, while the
remaining two need to be performed numerically.
There have been a number of attempts
\cite{dergachev_loosely_2012,haris_performance_2017,dhurandhar_marginalizing_2017,wette_geometric_2021}
to find usable solutions and approximations to these integrals, with the most successful
fully-analytic approximation to date provided by Bero~\&~Whelan \cite{2019CQGra..36a5013B}.

On the other hand, recent work \cite{covas_constraints_2022} on semi-coherent all-sky searches for
neutron stars in unknown binary systems has revealed some unexpected weaknesses of the
$\F$-statistic:
the extreme computational cost of these searches requires using very short segments, of order
$\Ord{\SI{100}{\second}}$.
Surprisingly, the $\F$-statistics turns out to be quite sub-optimal for such short segments, and an
empirically-constructed \emph{dominant-response} statistic $\F\sub{AB}$, effectively dropping two of
the four amplitude degrees of freedom, was shown in \cite{covas_improved_2022} to beat the
$\F$-statistic sensitivity by up to $\sim\SI{19}{\percent}$ in the short-segment limit.
Furthermore, when signal power varies over segments, which can happen both due to varying
(i) antenna-pattern response, and (ii) noise floors and duty factors, then an explicit
segment-weighting scheme was shown in \cite{covas_improved_allsky_2022} to improve sensitivity over
the conventional directly-summed semi-coherent $\sco{\F}$- (or $\sco{\F}\sub{AB}$) statistics.

These empirical findings raise the question of how to understand and leverage them within the
Bayesian framework.
Given that three of the four amplitude parameters (namely $\iota,\psi,\phiO$) have well-defined
ignorance priors \cite{2009CQGra..26t4013P}, the only remaining freedom left is the choice of
$\hO$-prior.  Previous works on the CW Bayes factor have all used a uniform $\hO$-prior, for
simplicity and because it enables analytical marginalization.
However, this is intrinsically a ``strong-signal'' prior, in the sense that it puts more probability
weight on larger orders of magnitude compared to smaller ones.

Here we explore a new $\hO$-prior in the form of a half-Gaussian with scale parameter $H$, which is
more ``physical'' in the sense of giving more weight to smaller $\hO$ amplitudes compared to larger
ones.
This contains the uniform prior in the asymptotic strong-signal limit ($H\rightarrow\infty$), while
preserving the ability for analytic $\hO$ integration.
In the ``weak-signal'' limit ($H\rightarrow 0$), we can Taylor-expand in small $H$ and obtain a
fully analytic solution for all four Bayes- factor integrals.
Used as a semi-coherent statistic, this performs as well and better than the weighted
$\sco{\F}\sub{ABw}$ statistic for short segments and the weighted $\F$-statistic $\sco{\F}\sub{w}$
for long segments.

The plan of this paper is as follows: introduction to the $\F$-statistic formalism in
Sec.~\ref{sec:f-stat-cont}, followed by a recap of the corresponding Bayesian formalism in
Sec.~\ref{sec:gener-b-stat} with $\B$-statistic and Bero\&Whelan approximation.  Derivation of the
new generalized $\B$-statistic and analytic weak-signal approximation in
Sec.~\ref{sec:gener-b-stat-weak}.  Derivation of the weak-signal statistic as a generalized
$\chi^2$-distribution in Sec.~\ref{sec:distr-weak-sign-1}, followed by numerical tests and results
presented in Sec.~\ref{sec:numerical-results}, with conclusions in Sec.~\ref{sec:conclusions}.
\ref{sec:altern-deriv-weak} shows an alternative way to derive the weak-signal statistic,
\ref{sec:weight-multi-detect} gives the expressions for the general multi-detector
non-stationary scalar product, \ref{sec:scal-prod-phys} provides the explicit likelihood terms
in physical coordinates, and \ref{sec:relation-5-vector} shows the detailed translation to the
5-vector framework and weak-signal statistic equivalence.

\section{The $\F$-statistic formalism}
\label{sec:f-stat-cont}

\subsection{Signal model}
\label{sec:signal-model}

The signal depends on amplitude parameters $\Amp \equiv \{\hO,\cosi,\psi,\phiO\}$, where $\hO$ is
the overall amplitude, $\cosi\equiv\cos\iota$ quantifies the degree of circular polarization in
terms of the inclination angle $\iota$ of the neutron-star spin axis to the line of sight, $\psi$
describes the polarization in terms of its rotation angle on the sky, and $\phiO$ is the signal
phase at a reference time (see \cite{2014CQGra..31f5002W} for a more detailed discussion of the
geometry). We refer to this parametrization as the \emph{physical} amplitude coordinates.

The seminal paper \cite{jks98:_data} introduced a new set of amplitude coordinates
$\{\Amp^\mu\}$, referred to as JKS coordinates, which are defined as
\begin{equation}
  \label{eq:27}
  \eqalign{
    \Amp^{1} &\equiv \hspace{0.7em}\Aplus \cos \phiO \cos 2 \psi - \Across \sin \phiO \sin 2 \psi\,, \\
    \Amp^{2} &\equiv \hspace{0.7em}\Aplus \cos \phiO \sin 2 \psi + \Across \sin \phiO \cos 2 \psi\,, \\
    \Amp^{3} &\equiv -\Aplus \sin \phiO \cos 2 \psi - \Across \cos \phiO \sin 2 \psi\,, \\
    \Amp^{4} &\equiv -\Aplus \sin \phiO \sin 2 \psi + \Across \cos \phiO \cos 2 \psi\,,
  }
\end{equation}
in terms of the polarization amplitudes $\Aplus \equiv \hO\,( 1 + \cosi^2)/2$ and $\Across \equiv \hO\,\cosi$.
Using these coordinates, the signal $h(t)$ in the detector frame takes the linear form
\begin{equation}
  \label{eq:49}
  h(t;\Amp,\dop) = \Amp^\mu\, h_{\mu} (t; \dop),
\end{equation}
with automatic summation over repeated amplitude indices  $\mu=1,\ldots,4$,
and with four matched-filter basis functions $h_\mu(t;\dop)$ defined as
 \begin{equation}
  \label{eq:basisfunc}
  \eqalign{
    h_{1}(t)&\equiv a(t) \, \cos \phi(t),\quad h_{2}(t) \equiv b(t) \,\cos \phi(t), \\
    h_{3}(t)&\equiv a(t) \, \sin \phi(t), \quad  h_{4}(t) \equiv b(t)\,\sin \phi(t),
  }
\end{equation}
in terms of the signal phase $\phi(t;\dop)$ at the detector with sky-direction dependent
antenna-pattern functions $a(t;\sky),\,b(t;\sky)$ (cf.\ \cite{jks98:_data,prix:_cfsv2} for explicit
expressions).

The \emph{phase-evolution parameters} $\dop$ include the remaining signal parameters, namely the
sky-position $\sky$, frequency $f$ and higher-order derivatives $\dot{f},\ddot{f},\ldots$ at some
given reference time, describing the slowly-changing intrinsic signal frequency.  For CW sources in
binary systems $\dop$ also includes the binary orbital parameters.  The sky position $\sky$ is most
commonly specified in terms of the equatorial longitude $\alpha$ and latitude $\delta$, measured in
radians.

\subsection{Likelihood ratio}
\label{sec:likelihood-ratio}

We want to construct the \emph{optimal} detection statistic for distinguishing between the
Gaussian-noise hypothesis $\HypG: x(t) = n(t)$, when the data $x$ contains only Gaussian noise $n$,
and the signal hypothesis $\HypS: x(t) = n(t) + h(t;\Amp,\dop)$, when there is an additional signal
$h(t;\Amp,\dop)$ of the form \eref{eq:49}.
\emph{Optimality} of a statistic $\stat(x)$ is defined in terms of the Neyman-Pearson criterion of
maximizing the detection probability $p\DET=\prob{\stat>\stat\thresh}{\HypS}$ at a fixed false-alarm
probability $p\FA \equiv\prob{\stat>\stat\thresh}{\HypG}$, parametrized in terms of a detection
threshold $\stat\thresh$.

For two \emph{simple} hypothesis, i.e., with no unknown parameters, the classic Neyman-Pearson lemma
\cite{stuart_kendalls_1999} proves that the optimal statistic is the likelihood ratio, namely
\begin{equation}
  \label{eq:18}
  \Lr(x;\Amp,\dop) \equiv \frac{\prob{x}{\HypS,\Amp,\dop}}{\prob{x}{\HypG}}
  = \exp\left[\scalar{x}{h} - \frac{1}{2}\scalar{h}{h}\right],
\end{equation}
where the last expression holds for the specific hypotheses $\HypG,\HypS$ being considered here, and
we introduced the ``matched-filter'' scalar product $\scalar{x}{y}$, see \cite{finn_detection_1992}.

The scalar product for narrow-band signals (such as CWs) in a single detector with stationary noise
floor and no gaps in the data, would simply be
$\scalar{x}{y} \equiv 2\Sn\inv\int_0^{\Tdata} x(t)\,y(t)\,dt$,
in terms of the (single-sided) noise power spectral density $\Sn(f)$ around the signal frequency $f$
and data duration $\Tdata$.
One can generalize this to several detectors, allowing for gaps in the data and non-stationary
noise-floors (see \ref{sec:weight-multi-detect} for the full expression), and still write it in
the form
\begin{equation}
  \label{eq:29}
  \scalar{x}{y} \equiv 2\dfact\,\av{x\,y}\,,
\end{equation}
in terms of a (weighted) multi-detector average $\av{\ldots}$ and a dimensionless \emph{data factor}
$\dfact$ \cite{prix:_cfsv2,covas_improved_2022}, which we define as
\begin{equation}
  \label{eq:30}
  \dfact \equiv \Sn\inv\, \Tdata,
\end{equation}
which can be understood as the product of the data ``quality'' (i.e., $\Sn\inv$) and ``quantity'',
$\Tdata$.
Note that this involves the non-stationary generalization of the overall noise floor $\Sn$ given in
\eqref{eq:32}.

Substituting the JKS signal form \eqref{eq:49}, we can express the log-likelihood ratio as
\begin{equation}
  \label{eq:35}
  \ln\Lr(x;\Amp,\dop) = \Amp^\mu x_\mu - \frac{1}{2}\Amp^\mu\M_{\mu\nu} \Amp^\nu\,,
\end{equation}
where we defined
\begin{equation}
  \label{eq:41}
  x_\mu \equiv \scalar{x}{h_\mu},\quad\text{and}\quad
  \M_{\mu\nu}\equiv \scalar{h_\mu}{h_\nu}.
\end{equation}
The four numbers $x_\mu$ are the scalar products of the data $x$ matched against the basis functions
$h_\mu$ of \eqref{eq:basisfunc}, while the \emph{detector-response matrix} $\M_{\mu\nu}$ can be
expressed more explicitly\footnote{Assuming the long-wavelength limit for ground-based
  detectors.} as
\begin{equation}
  \label{eq:74}
  \M = \dfact M = \dfact \left(\begin{array}{cc}
    m & 0\\
    0 & m\\
  \end{array}\right),
\quad\text{with}\quad
  m \equiv
  \left(\begin{array}{cc}
    A &  C \\
    C &  B \\
  \end{array}
\right),
\end{equation}
where $A=\av{a^2}$, $B=\av{b^2}$, and $C=\av{a\,b}$ are the (sky-position
dependent) averaged antenna-pattern coefficients.
This expression shows two contributions to the detector response, namely the data factor $\dfact$
and a geometric \emph{antenna-pattern matrix} $M$, encoding the detector sensitivity to a particular
sky direction (integrated over the available observation time).
We can express the determinant as
\begin{equation}
  \label{eq:46}
  \det\M = \dfact^4\,D^2\,\quad\text{with}\quad
  D \equiv \det m = A B - C^2\,.
\end{equation}

In practice the amplitude parameters $\Amp$ are (generally) unknown, and depending on the
type of search, also some or all of the phase-evolution parameters.
The signal hypothesis $\HypS$ is therefore always a \emph{composite} hypothesis involving unknown
parameters.
Therefore the likelihood ratio \eqref{eq:18} is a function of these unknown
parameters and cannot be directly used as a detection statistic.

\subsection{The $\F$-statistic}
\label{sec:f-statistic-bayes}

A classic method for dealing with composite hypotheses is the maximum-likelihood approach, which
consists of using the maximized likelihood ratio as a detection statistic.  Applying this approach
to \eqref{eq:35}, one can analytically maximize $\Lr$ over the JKS amplitude parameters $\Amp^\mu$
and obtain
\begin{equation}
  \label{eq:31}
  \F(x;\dop) \equiv \max_{\{\Amp^\mu\}} \ln\Lr(x;\Amp,\dop) = \frac{1}{2} x_\mu \left(\M\inv\right)^{\mu\nu}x_\nu,
\end{equation}
defining the $\F$-statistic \cite{jks98:_data}, with the corresponding maximum-likelihood estimators
$\mle{\Amp}^\mu$ for the amplitude parameters given by
\begin{equation}
  \label{eq:25}
  \mle{\Amp}^\mu \equiv \left(\M^{-1}\right)^{\mu\nu}x_\nu\,.
\end{equation}
This statistic can be shown \cite{jks98:_data} to follow a $\chi^2$-distribution with \emph{four}
degrees of freedom, and noncentrality parameter
\begin{equation}
  \label{eq:45}
  \rho^2 \equiv \scalar{h}{h}= \Amp^\mu\M_{\mu\nu}\Amp^\nu,
\end{equation}
which we refer to as the \emph{signal power}, also commonly known as the squared signal-to-noise
ratio (SNR) in this context.

\section{Bayesian detection framework}
\label{sec:gener-b-stat}

\subsection{Bayes factor }
\label{sec:bayes-factor-}

Unlike the \emph{ad-hoc} maximum-likelihood approach discussed earlier, the Bayesian framework
offers a natural and unique method for handling unknown parameters, grounded in the three
fundamental laws of probability \cite{jaynes_probability_2003}.

The likelihood for a composite signal hypothesis $\HypS$ can be directly obtained as
\begin{equation}
  \label{eq:33}
  \prob{x}{\HypS} = \int \prob{x}{\HypS,\Amp,\dop}\,\prob{\Amp,\dop}{\HypS}\,d\Amp\,d\dop,
\end{equation}
also referred to as the \emph{marginalized} likelihood, which requires an explicit specification of
the prior probabilities for the unknown signal parameters $\{\Amp,\dop\}$.
Assuming for simplicity a Gaussian-noise hypothesis with known noise floor $\Sn$, this further leads
to the marginalized likelihood ratio
\begin{equation}
  \label{eq:39}
  B\sub{\Signal/\Gauss}(x) \equiv \frac{\prob{x}{\HypS}}{\prob{x}{\HypG}}
  = \int \Lr(x;\Amp,\dop)\,\prob{\Amp,\dop}{\HypS}\,d\Amp\,d\dop,
\end{equation}
commonly known as the \emph{Bayes factor} between the signal and the Gaussian-noise hypothesis.

Here we focus exclusively on the marginalization over amplitude parameters $\Amp$, and therefore
consider only a single phase-evolution point $\dop$ in the following.
This could be interpreted as a ``targeted'' search scenario, but could equally represent simply a
single template step in a wide-parameter space search.

\subsection{Rediscovering the $\F$-statistic}
\label{sec:recov-f-stat}

Building on the pioneering work \cite{searle_robust_2008,searle_bayesian_2009} in burst
searches, \cite{2009CQGra..26t4013P} demonstrated that the $\F$-statistic can also be derived as
a Bayes factor, assuming a (somewhat arbitrary) prior that is uniform in $\Amp^\mu$-space.
Specifically, $\prob{\{\Amp^\mu\}}{\HypS}=\AmuPriorConst$, leading to a Gaussian integral that yields
\begin{equation}
  \label{eq:44}
  B\sub{\Signal/\Gauss}(x;\dop) = \AmuPriorConst\frac{4\pi^2}{\sqrt{\det\M}} e^{\F(x;\dop)}\,.
\end{equation}
The Jacobian for the coordinate transformation \eqref{eq:27} between $\Amp^\mu$ and physical
amplitude parameters can be obtained \cite{2009CQGra..26t4013P,2014CQGra..31f5002W} as
\begin{equation}
  \label{eq:11}
  \prob{\hO,\cosi,\psi,\phiO}{\HypS} = \frac{1}{4}\hO^3\left(1-\cosi^2\right)^3\prob{\{\Amp^\mu\}}{\HypS}\,,
\end{equation}
therefore a constant-$\Amp^\mu$ prior favors signals with large $\hO$ and linear polarization
($\cosi\sim0$) over circular polarization $\cosi\sim\pm 1$, both of which are unphysical.

In order to be normalizable, this prior needs to be artificially truncated at some large-$\Amp^\mu$
surface. One can also use this to (arbitrarily) remove the prefactor by choosing an $\M$-dependent
truncation surface, as discussed in \cite{prix_search_2011}, resulting in a $\F$-statistic Bayes
factor of the form $B\sub{\Signal/\Gauss}(x;\dop) = \AmuPriorConst'\,e^{\F(x;\dop)}$ instead, where
the prefactor $\AmuPriorConst'$ is now independent of the detector-response matrix $\M$.
Empirically this form was found to result in better performance on transient-CW signals
\cite{prix_search_2011}, and was also used as the basis for constructing extended ``line-robust''
Bayes factors in \cite{keitel_search_2014,keitel_robust_2016}.

\subsection{Neyman-Pearson-Searle optimality}
\label{sec:neyman-pearson-searle}

As Searle noted in \cite{searle_monte-carlo_2008}, the Neyman-Pearson proof for the optimal
detection statistic only requires the likelihoods for the two competing hypotheses. Since the
Bayesian framework uniquely provides a marginalized likelihood \eqref{eq:33} for the composite
signal hypothesis, it follows that the marginalized likelihood ratio \eqref{eq:35} (i.e., the Bayes
factor) \emph{is} the Neyman-Pearson optimal statistic, assuming the unknown signal parameters are
drawn from the priors.

Thus, the somewhat unphysical prior of Sec.~\ref{sec:recov-f-stat} implies that the $\F$-statistic
is not optimal, and using more appropriate priors can lead to a more powerful detection
statistic. This was demonstrated in \cite{2009CQGra..26t4013P}, where an isotropic prior on the
neutron star's spin orientation (reflecting our ignorance of the spin axis) was used instead,
namely
\begin{equation}
  \label{eq:23}
  \prob{\hO,\cosi,\psi,\phiO}{\HypS} = \frac{1}{2\pi^2}\,\prob{\hO}{\HypS}\,,
\end{equation}
which is uniform in $\cosi\in[-1,1]$, $\psi\in[-\pi/4,\pi/4]$ and $\phiO\in[0, 2\pi]$.
We can therefore write the general form of the optimal Bayes factor as
\begin{equation}
  \label{eq:168}
  B\sub{\Signal/\Gauss}(x;\dop) \equiv \frac{1}{2\pi^2}
    \int \Lr(x;\dop,\Amp)\,\prior{\hO}\,d \hO\,d\cosi\,d\psi\,d\phiO\,.
\end{equation}
The correct choice for an $\hO$-prior is less obvious and would ultimately have to be constructed
from astrophysical arguments about the expected distance, rotation rate and deformation of neutron
stars.  Generally speaking, however, we expect a ``physical'' $\hO$-prior to favor weaker signals
over stronger ones, contrary to the implicit $\F$-statistic prior $\propto\hO^3$ of \eqref{eq:11}.

\subsection{Likelihood ratio in physical coordinates}
\label{sec:likel-ratio-phys}

In order to make progress on the Bayes-factor integral \eqref{eq:168}, it is useful to express the
likelihood ratio $\Lr$ of \eqref{eq:35} in terms of the physical amplitude parameters
$\{\hO,\cosi,\psi,\phiO\}$.
This can be obtained \cite{2014CQGra..31f5002W} in the following form,
\begin{equation}
  \label{eq:42}
  \ln\Lr(x;\Amp) = \hrel\, q(x;\cosi,\psi)\, \cos\left(\phiO - \varphi_0\right)
  - \frac{1}{2}\hrel^2\,\rhog^2(\cosi,\psi),
\end{equation}
where we defined the \emph{relative} signal amplitude $\hrel$ as
\begin{equation}
  \label{eq:63}
  \hrel \equiv \hO\sqrt{\dfact} = \hO \sqrt{\frac{\Tdata}{\Sn}}\,,
\end{equation}
and the geometric response function\footnote{This is closely related to the definition used in
\cite{wette2011:_sens,2018arXiv180802459D}, namely $R\equiv5\rhog/2$, chosen such that the
polarization- and sky average is $\av{R^2}_{\sky,\cosi,\psi}=1$.}
\begin{equation}
  \label{eq:47}
  \rhog^2(\cosi,\psi) \equiv \alpha_1\,A + \alpha_2\,B + 2\alpha_3\,C\,,
\end{equation}
in terms of the amplitude coefficients $\alpha_i(\cosi,\psi)$:
\begin{equation}
  \label{eq:72}
  \eqalign{
    \alpha_1(\cosi,\psi) &\equiv \frac{(\Amp^1)^2+(\Amp^3)^2}{\hO^2} = \frac{1}{4}(1+\cosi^2)^2\,\cos^22\psi + \cosi^2\,\sin^22\psi,\\
    \alpha_2(\cosi,\psi) &\equiv \frac{(\Amp^2)^2 + (\Amp^4)^2}{\hO^2} = \frac{1}{4}(1+\cosi^2)^2\,\sin^22\psi + \cosi^2\,\cos^22\psi,\\
    \alpha_3(\cosi,\psi) &\equiv \frac{\Amp^1\Amp^2 + \Amp^3\Amp^4}{\hO^2} = \frac{1}{4}(1-\cosi^2)^2\,\sin2\psi\cos2\psi.
  }
\end{equation}
This allows us to write the \emph{signal power} $\rho^2$ of (\ref{eq:45}) as
\begin{equation}
  \label{eq:81}
  \rho^2 = \hrel^2\,\rhog^2(\cosi,\psi)\,.
\end{equation}
The explicit form for the phase offset $\varphi_0(x;\cosi,\psi)$ in \eqref{eq:42} will not be
required in the following, but can be found in \eqref{eq:17}.
The matched-filter response term $q(x;\cosi,\psi)$ in \eqref{eq:42} can be expressed as
\begin{equation}
  \label{eq:100}
  q^2(x;\cosi,\psi) \equiv 2\F\sub{A}(x)\,\alpha_1 A + 2\F\sub{B}(x)\,\alpha_2 B
  + 4\F\sub{C}(x)\,\alpha_3 C\,,
\end{equation}
where we used the ``partial'' $\F$-statistics of \cite{covas_improved_2022}, namely
\begin{equation}
  \label{eq:40}
    2\F\sub{A}(x) \equiv \frac{x_1^2 + x_3^2}{\dfact A},\quad
    2\F\sub{B}(x) \equiv \frac{x_2^2 + x_4^2}{\dfact B},\quad
    2\F\sub{C}(x) \equiv \frac{x_1\,x_2 + x_3\,x_4}{\dfact C},
\end{equation}
which can be used to write the $\F$-statistic \eqref{eq:31} as
\begin{equation}
  \label{eq:71}
  2\F(x) = \frac{2}{D} \left[ A B ( \F\sub{A}(x) + \F\sub{B}(x) ) - 2C^2 \F\sub{C}(x)\right].
\end{equation}
For data $x=n+s$ containing a signal $s$, one can show that
\begin{equation}
  \label{eq:5}
  \eqalign{
    \expect{x_\mu} &= s_\mu\,\quad\text{with}\quad s_\mu\equiv \scalar{s}{h_\mu},\quad\text{and}\\
    \expect{x_\mu x_\nu} & =\M_{\mu\nu} + s_\mu s_\nu,
  }
\end{equation}
which allows one to show that
\begin{equation}
  \label{eq:101}
  \expect{2\F\sub{A}} = 2 + \rho\sub{A}^2,\quad
  \expect{2\F\sub{B}} = 2 + \rho\sub{B}^2,\quad
  \expect{2\F\sub{C}} = 2 + \rho\sub{C}^2\,,
\end{equation}
in terms of the respective ``partial'' signal powers
\begin{equation}
  \label{eq:102}
  \rho\sub{A}^2 \equiv \frac{s_1^2+s_3^2}{\dfact A},\quad
  \rho\sub{B}^2 \equiv \frac{s_2^2+s_4^2}{\dfact B},\quad
  \rho\sub{C}^2 \equiv \frac{s_1 s_2 + s_3 s_4}{\dfact C}\,.
\end{equation}
As shown in \cite{covas_improved_2022}, $2\F\sub{A}$ and $2\F\sub{B}$ are $\chi^2$-distributed
with two degrees of freedom and non-centrality parameters $\rho^2\sub{A}$ and $\rho^2\sub{B}$,
respectively.

\subsection{The $\B$-statistic}
\label{sec:bayes-fact-ampl}

With the likelihood ratio expressed in physical coordinates \eqref{eq:42}, the Bayes-factor integral
\eqref{eq:168} now takes the form
\begin{equation}
  \label{eq:181}
  B\sub{\Signal/\Gauss}(x) = \frac{1}{2\pi^2}\int d \hO d\cosi d\psi\,\prior{\hO}
    \,e^{-\frac{1}{2}\hrel^2\rhog^2}\int e^{\hrel q\cos(\phiO - \varphi_0)}\,d\phiO\,.
\end{equation}
The $\phiO$-integral can be performed analytically \cite{2014CQGra..31f5002W} using the Jacobi-Anger
expansion, resulting in $\int_{0}^{2\pi}e^{x\,\cos\phi}\,d\phi = 2\pi\,I_0(x)$ in terms of the
modified Bessel function of the first kind $I_0(x)$, which yields
\begin{equation}
  \label{eq:189}
  B\sub{\Signal/\Gauss}(x) = \frac{1}{\pi}\int
  \,e^{-\frac{1}{2}\hrel^2\rhog^2}\, I_0\left(\hrel q\right)\,
  \prior{\hO}\, d\hO\,d\cosi\,d\psi.
\end{equation}
In \cite{2009CQGra..26t4013P} and all subsequent analyses of this Bayes factor, only the uniform
$\hO$ prior, $\prior{\hO} = \hOPriorConst$ (for $\hO \in [0, h_{\max{}}]$), has been used.
This simple choice allows for analytic integration over $\hO$, serving as a proof of principle by
demonstrating that the resulting Bayes factor is more sensitive than the $\F$-statistic.
The prior leads to a known integral \cite{2014CQGra..31f5002W} for $\hO$, specifically using
Eq.~11.4.31 in Abramowitz~\&~Stegun \cite{abramowitz_handbook_1964}:
\begin{equation}
  \label{eq:192}
  \int_{0}^{\infty} e^{-a^2\,t^2}\,I_0(b\,t)\,dt = \frac{\pi^{\frac{1}{2}}}{2 a}\,e^{\frac{b^2}{8a^2}}\,I_0\left(\frac{b^2}{8a^2}\right)\,,
\end{equation}
and so we are left with a two-dimensional integral
\begin{equation}
  \label{eq:201}
  \eqalign{
    \B(x) &\equiv \frac{\hOPriorConst}{\pi} \int\frac{1}{\rhog}\,e^{\Theta}\,I_0(\Theta)\, d\cosi\,d\psi,\quad\text{with}\\
    \Theta(x;\cosi,\psi) &\equiv \frac{q^2(x;\cosi,\psi)}{4\rhog^2(\cosi,\psi)},
  }
\end{equation}
defining the ``$\B$-statistic''.

Several studies have investigated this statistic and its integral, exploring different amplitude
coordinate systems and testing various approaches to derive useful analytical approximations
\cite{dergachev_loosely_2012,2014CQGra..31f5002W,haris_performance_2017,dhurandhar_marginalizing_2017,2019CQGra..36a5013B}.
Recently, a novel geometric approach to expressing the likelihood ratio and deriving the
$\F$-statistic through marginalization was explored in \cite{wette_geometric_2021}.
In the next section we will discuss the approach of \cite{2019CQGra..36a5013B}, which
is the most effective analytic approximation to \eqref{eq:201} found so far.

\subsection{The Bero-Whelan approximation $\B\BeroWhelan(x)$}
\label{sec:bero-whel-appr}

This approximation is expressed using circular-polarization-factored (CPF) amplitude coordinates
$\Amp^{\cpf{\mu}}$, introduced in \cite{2014CQGra..31f5002W} as the linear combinations
\begin{equation}
  \label{eq:15}
  \eqalign{
    \Amp^{\cpf{1}}\equiv\frac{1}{2}\left(\Amp^1 + \Amp^4\right),&\quad
    \Amp^{\cpf{2}}\equiv\frac{1}{2}\left(\Amp^2 - \Amp^3\right)\,,\\
    \Amp^{\cpf{3}}\equiv\frac{1}{2}\left(\Amp^1 - \Amp^4\right),&\quad
    \Amp^{\cpf{4}}\equiv\frac{1}{2}\left(-\Amp^2 - \Amp^3\right)\,,
  }
\end{equation}
with corresponding transformation of the detector-response matrix \eqref{eq:74}
into\footnote{This definition differs from \cite{2014CQGra..31f5002W,2019CQGra..36a5013B} by
  factoring out the data factor $\dfact$.}
\begin{equation}
  \label{eq:19}
  \M_{\cpf{\mu}\cpf{\nu}}=\dfact
  \left(\begin{array}{cccc}
    I & 0 & L & -K \\
    0 & I & K & L  \\
    L & K & J & 0  \\
    -K& L & 0 & J
  \end{array}\right),
\end{equation}
in terms of the CPF antenna-pattern coefficients
\begin{equation}
  I = J = A + B,\quad K = 2C,\quad L = A - B\,,
\end{equation}
where we assumed the long-wavelength limit.

The Bayes-factor approximation follows from assuming small off-diagonal terms $K/I=2C/(A+B)$ and
$L/I=(A-B)/(A+B)$, and Taylor-expanding to first order around a diagonal CPF antenna-pattern matrix,
which eventually can be shown to result in the explicit solution
\begin{eqnarray}
  \fl\ln\frac{\B\BeroWhelan(x)}{\B(0)} \equiv \ln b_0(\mle{y}\sub{R}) + \ln b_0(\mle{y}\sub{L})
  + \mle{A}\sub{R}\mle{A}\sub{L}\dfact\left[K\sin4\mle{\psi} + L\cos 4\mle{\psi}\right]\nonumber\\
  \times \left[
  \frac{1}{4}\frac{b_1(\mle{y}\sub{R})}{b_0(\mle{y}\sub{R})}
  + \frac{1}{4}\frac{b_1(\mle{y}\sub{L})}{b_0(\mle{y}\sub{L})}
  - \frac{1}{16}\frac{b_1(\mle{y}\sub{R})}{b_0(\mle{y}\sub{R})}\frac{b_1(\mle{y}\sub{L})}{b_0(\mle{y}\sub{L})}
  \right],
  \label{eq:20}
\end{eqnarray}
in terms of the functions
\begin{equation}
  \label{eq:21}
  \eqalign{
    \ln b_0(y) \equiv \ln {}_1F_1\left(\frac{1}{4}, 1, y\right)\stackrel{y\gg1}{\sim}
    -\ln\Gamma\left(\frac{1}{4}\right) + y - \frac{3}{4}\ln y,\\
    \ln b_1(y) \equiv \ln {}_1F_1\left(\frac{5}{4}, 2, y\right)\stackrel{y\gg1}{\sim}
    -\ln\Gamma\left(\frac{5}{4}\right) + y - \frac{3}{4}\ln y,
  }
\end{equation}
where the asymptotic forms \cite{2019CQGra..36a5013B,asympt_confluent_hypergeom} are used for
$y>700$ to avoid numerical overflow.
We further defined the shortcuts
\begin{equation}
  \label{eq:22}
  \mle{y}\sub{R} \equiv \frac{\dfact I \mle{A}\sub{R}^2}{2},\quad
  \mle{y}\sub{L} \equiv \frac{\dfact J \mle{A}\sub{L}^2}{2}\,,
\end{equation}
and the above expressions are evaluated at the maximum-likelihood amplitude estimators
$\mle{\Amp}^\mu$ of \eqref{eq:25}, translated to polar CPF coordinates \cite{2014CQGra..31f5002W}
\begin{equation}
  \label{eq:26}
  A\sub{R} = \frac{\Aplus + \Across}{2},\quad A\sub{L}=\frac{\Aplus - \Across}{2},
\end{equation}
and polarization angle $\psi$, which can be obtained by inverting \eqref{eq:27}.
The normalization $\B(0)\approx \frac{\kappa \Gamma(1/4)^2}{2^{5/2}(\dfact^2IJ)^{1/4}}$ represents a
(sky-dependent) offset and does not affect the detection power of the statistic.

\section{Generalized $\B$-statistic and weak-signal approximation}
\label{sec:gener-b-stat-weak}

\subsection{Motivation}
\label{sec:motivation}

Recent work \cite{covas_improved_2022} has revealed some surprising shortcomings of the
$\F$-statistic for short coherence times $T\lesssim \SI{1}{day}$. In fact, it was explicitly shown
that an empirically-constructed ``dominant-response'' statistic $\F\sub{AB}$, defined as
\begin{equation}
  \label{eq:60}
  \F\sub{AB}(x) \equiv \left\{\F\sub{A}(x)\quad\text{if}\;A\ge B,\quad \F\sub{B}(x)\;\text{otherwise}\right\},
\end{equation}
using the partial $\F$-statistics $\F\sub{A}$ and $\F\sub{B}$ of \eqref{eq:40}, can
\emph{substantially} improve upon the sensitivity of the $\F$-statistic for short coherence times,
by up to $\SI{19}{\percent}$.

In the Bayesian framework there is no handle for encoding ``short segments'', and in fact the only
remaining freedom in the $\B$-statistic definition \eqref{eq:189} is in the choice of $\hO$-prior, $\prior{\hO}$.
However, considering that shorter coherence times will correspond to less signal power (\ref{eq:81}),
motivates the idea of exploring an $\hO$-prior that allows for describing ``weak signals''.

\subsection{New $\hO$ prior: introducing an amplitude scale $H$}
\label{sec:gener-b-stat-prior}

We consider a half-Gaussian prior on $\hO$ with scale parameter $H$, namely
\begin{equation}
  \label{eq:190}
  \prior{\hO} = \frac{\sqrt{2}}{\sqrt{\pi}\,H}\,
  e^{-\frac{\hO^2}{2H^2}},\quad\text{for }\hO\ge 0.
\end{equation}
This prior preserves the functional form of the integrand \eqref{eq:189} and therefore still allows
for analytic $\hO$ integration via \eqref{eq:192} as with the uniform prior, resulting in the
generalized $\B\sub{H}$-statistic:
\begin{equation}
  \label{eq:57}
  \eqalign{
    \B\sub{H}(x) &\equiv \frac{1}{\pi}\int d\cosi\,d\psi
    \frac{1}{\sqrt{1 + \Hrel^2\rhog^2}}\,e^{\Theta\sub{H}}\, I_0(\Theta\sub{H})\,,\quad\text{with}\\
    \Theta\sub{H}(x;\cosi,\psi) &\equiv \frac{\Hrel^2 \,q^2}{4(1 + \Hrel^2\,\rhog^2)}\,,
  }
\end{equation}
where we defined the \emph{relative} scale $\Hrel$ in analogy to \eqref{eq:63}, as
\begin{equation}
  \label{eq:103}
  \Hrel \equiv H\,\sqrt{\dfact}\,.
\end{equation}
In the strong-signal limit ($H\rightarrow\infty$) this prior converges to a uniform prior, and we
can also recover the standard $\B$-statistic of \eqref{eq:201} in that limit, namely
$\B\sub{H}(x) \overset{H\rightarrow \infty}{\longrightarrow} \B(x)$.
Furthermore, the half-Gaussian prior is more ``physical'' than the uniform one in two respects:
(i) it favors weaker signals over stronger ones, as would be physically expected, and
(ii) it is normalizable and therefore does not rely on an arbitrary cutoff at some large $\hO$ value.

From a practical point of view, the new prior does not help with analytically solving the remaining
two marginalization integrals, and is therefore as computationally impractical as the
$\B$-statistic.
It does, however, open up a new limiting regime of \emph{weak signals}, when $H\rightarrow 0$, which
we will explore next.

\subsection{Weak-signal approximation $\B\weak(x)$}
\label{sec:weak-signal-bayes}

Assuming the weak-signal limit $H\rightarrow0$ of the new $\hO$-prior \eqref{eq:190}, we can
Taylor-expand the integrand \eqref{eq:57} in terms of small $\Hrel\ll 1$, namely
\begin{equation}
  \label{eq:65}
  \eqalign{
    \frac{1}{\sqrt{1 + \Hrel^2\rhog^2}} &= 1
    -\frac{1}{2}\Hrel^2\rhog^2 + \Ord{\Hrel^4},\\
    \Theta\sub{H}(x;\cosi,\psi) &= \frac{1}{4}\Hrel^2 q^2 + \Ord{\Hrel^4},\\
    e^{\Theta\sub{H}} &= 1 + \frac{1}{4}\Hrel^2 q^2 + \Ord{\Hrel^4},\\
    I_0(\Theta\sub{H}) &= 1 + \Ord{\Hrel^4},
  }
\end{equation}
and to leading order in $\Hrel$ the $\B\sub{H}$-statistic now takes the form
\begin{equation}
  \label{eq:66}
  \eqalign{
    \B\sub{H}(x) &= \frac{1}{\pi}\int d\cosi\,d\psi \left( 1
      + \frac{1}{4}\Hrel^2\left(q^2 - 2 \rhog^2\right)\right) + \Ord{\Hrel^4}\\
            &= 1 + \frac{\Hrel^2}{4}\av{q^2 - 2\rhog^2}_{\cosi,\psi} + \Ord{\Hrel^4}.
  }
\end{equation}
Using the known $\cosi,\psi$-averages of \eqref{eq:72}, namely
$\av{\alpha_1}_{\cosi,\psi} = \av{\alpha_2}_{\cosi,\psi}=\frac{2}{5}$ and $\av{\alpha_3}_{\cosi,\psi}=0$,
applied to \eqref{eq:47} and \eqref{eq:100}, we find
\begin{equation}
  \label{eq:67}
    \av{\rhog^2}_{\cosi,\psi} = \frac{2}{5}\left(A+B\right),\quad
    \av{q^2}_{\cosi,\psi} =  \frac{2}{5}\left( 2\F\sub{A}(x)\,A + 2\F\sub{B}(x)\,B\right),
\end{equation}
resulting in the weak-signal Bayes-factor $\B\weak$ defined as
\begin{equation}
  \label{eq:68}
  \eqalign{
    \ln \B\weak(x) &\equiv \frac{H^2}{10}\left[ \dfact A\,\left(2\F\sub{A}(x) - 2\right) + \dfact B\,\left(2\F\sub{B}(x) - 2\right)\right]\\
    &= \frac{H^2}{10}\left[ \vx^2 - \trace\M\right]\,,
              }
\end{equation}
where boldface denotes a (column-) four-vector in
amplitude space, i.e., $\{\vx\}_\mu = x_\mu$, and $\vx^2\equiv\vx\transpose\vx=\sum_\mu x_\mu^2$.
It will be useful to define a simpler but equivalent $\bcoh$-statistic as
\begin{equation}
  \label{eq:124}
  \eqalign{
    \bcoh(\vx) &\equiv \frac{\vx^2}{\dfact} = A\,2\F\sub{A}(\vx) + B\,2\F\sub{B}(\vx),
  }
\end{equation}
which is of order $\bcoh\sim\Ord{1}$ in the absence of a signal and is independent of the prior
scale $H$.
It is interesting to compare this to the $\F$-statistic expressions of \eqref{eq:71} and
\eqref{eq:31}, and we can further also write it as
\begin{equation}
  \label{eq:2}
  2\F(x) = \frac{\vx\transpose M^{-1}\vx}{\dfact}.
\end{equation}
In terms of the $\bcoh$-statistic, the weak-signal Bayes factor \eqref{eq:68} is now just 
\begin{equation}
  \label{eq:1}
  \ln\B\weak(x) = \frac{\Hrel^2}{10}\,\left(\bcoh(\vx) - \trace M\right)\,,
\end{equation}
which is a monotonic function of $\bcoh$ and therefore an equivalent statistic.

We further point out that the coherent $\bcoh$-statistic is equivalent to the (initial)
5-vector-method statistic constructed in \cite{astone_method_2010}, which is further analyzed (in
terms of its noise distribution) in \cite{astone_method_2014}.  As noted in
\cite{astone_method_2010} and more recently discussed in greater detail in \cite{donofrio_two_2024},
a different choice of weights in that framework can also lead to the $\F$-statistic instead.
The details of the translation of the notation and formalism to see this equivalence are given in
\ref{sec:relation-5-vector}.

\subsection{Semi-coherent generalization}
\label{sec:semi-coher-gener}

A semi-coherent Bayes factor can be derived \cite{prix_search_2011} by relaxing the
signal hypothesis to allow for independent amplitude parameters $\Amp_\ell$ for ever segment
$\ell=1,\ldots,N\seg$.
The resulting semi-coherent Bayes factor is then given by the product of per-segment coherent Bayes
factors, each marginalized over its per-segment amplitude parameters $\Amp_\ell$ independently.
%
% It is interesting to note that in the weak-signal limit \eqref{eq:66} we obtain the same result
% assuming only independent $\{\hO,\phiO\}$ for each segment, but to $\Ord{\Hrel^2}$ one can still
% perform the $\{\cosi,\psi\}$ integration globally over all segments.
%
The resulting semi-coherent weak-signal Bayes factor is therefore
\begin{equation}
  \label{eq:36}
  \eqalign{
    \ln\sco{\B}\weak(x) &\equiv \sum_{\ell=1}^{N\seg} \ln\B\weak{}_{,\ell}(x)\\
    & = \frac{H^2}{10} \sum_\ell \left[
      \dfact_\ell A_\ell \left(2\F\sub{A,\ell}(x)-2 \right)
      + \dfact_\ell B_\ell\left(2\F\sub{B,\ell}(x)-2\right)
    \right].
  }
\end{equation}
This expression shows that each $\F\sub{A,\ell}(x)$ and $\F\sub{B,\ell}(x)$ contribution is
naturally summed over segments with respective \emph{segment weights} $(\dfact A)_\ell$ and
$(\dfact B)_\ell$, which account for varying data-factors (i.e., different quality and/or quantity
of available data) and antenna-pattern sensitivity (to a particular sky-direction $\vn$) over
segments.
Only recently \cite{covas_improved_allsky_2022} has shown that applying segment weights of the form
$\propto \dfact(A+B)$ to the $\F$-statistic (and similarly for the dominant-response statistic
$\F\sub{AB}$) improves their detection power, defining the weighted semi-coherent statistics as:
\begin{equation}
  \label{eq:43}
  \eqalign{
    \sco{\F}\sub{w}(x) &\equiv \sum_\ell w'_\ell\,\F_\ell(x)\,,\quad\text{with}\quad w'_\ell\equiv K'\dfact_\ell(A_\ell+B_\ell)\,,\\
    \sco{\F}\sub{AB,w}(x) &\equiv \sum_\ell w''_\ell\,\F\sub{AB,\ell}(x)\quad\text{with}\quad
    w''_\ell\equiv K''\dfact_\ell\left(Q_\ell+\frac{C_\ell}{Q_\ell}\right),
  }
\end{equation}
where $Q\equiv\max(A,B)$ and  $K',K''$ are weight normalizations, typically chosen to obtain unit
mean over segments, i.e., $\av{w}_\ell=1$.

Defining per-segment \emph{data weights} as
\begin{equation}
  \label{eq:3}
  w^\ell \equiv \frac{\dfact_\ell}{\mean{\dfact}},\quad\text{with}\quad
  \mean{\dfact} \equiv \frac{1}{N\seg}\sum_\ell \dfact_\ell\,,
\end{equation}
in terms of the average data factor $\mean{\dfact}$ over segments, such that
$\sum_\ell w^\ell = N\seg$, we can introduce a semi-coherent $\bsco$-statistic as
\begin{equation}
  \label{eq:125}
  \bsco(\vx) \equiv \sum_\ell w^\ell\, \bcoh_\ell(\vx) = \sum_\ell \frac{\vx_\ell^2}{\mean{\dfact}},
\end{equation}
and with a corresponding average relative scale defined as $\mean{\Hrel^2}\equiv H^2\mean{\dfact}$,
the semi-coherent Bayes factor now takes the form
\begin{equation}
  \label{eq:126}
  \ln\sco{\B}\weak(x) = \frac{\mean{\Hrel^2}}{10}\,\left( \bsco(\vx) - \trace\sco{M}\right),
\end{equation}
in terms of the semi-coherent antenna-pattern matrix
\begin{equation}
  \label{eq:4}
  \sco{M} = \sum_\ell w^\ell M_\ell = \frac{\sum_\ell \M_\ell}{\mean{\dfact}}.
\end{equation}

\subsection{Do detectable signals falls into the weak-signal regime?}
\label{sec:do-detect-sign}

Using the simple (albeit slightly biased \cite{wette2011:_sens,2018arXiv180802459D}) sensitivity
estimate of \cite{ligo_scientific_collaboration_setting_2004}, we can express the weakest detectable
signal amplitude as $\hO'\sim 11.4 \sqrt{\Sn / \Tdata}$ at a false-alarm probability of $p\FA=1\%$
and a detection probability of $p\DET=90\%$.  This corresponds to a relative amplitude of
$\hrel' = 11.4$ for a typical well-detectable signal.
Even for semi-coherent searches, assuming a rough $N\seg^{1/4}$ sensitivity scaling in terms of the
number of segments, this would only reach the weak-signal threshold of $\hrel\sim 1$ over a single
segment at around $N\seg\sim\Ord{\num{e4}}$.

This shows that for most typical searches, detectable signals would not actually be expected to fall
into the ``weak signal'' regime $\hrel\ll 1$.  On the other hand, the half-Gaussian prior has no
hard cutoff and therefore also covers strong signals.  We also see that for the resulting $\B\weak$
and $\bcoh$ statistics, increased signal power always translates into higher detection-statistic
values.

The weak-signal prior is sensible, however, exactly because any detectable CW signal is expected to
come from the tail of the actual $\hO$ distribution of existing astrophysical CW signals in the
data, most of which are in fact undetectably weak.

\section{Distribution of the $\beta$-statistic}
\label{sec:distr-weak-sign-1}

\subsection{Decorrelating $\{x_\mu\}$}
\label{sec:decorrelating-x_mu}

We see in \eqref{eq:68} that $\bcoh(\vx)$ is a quadratic function of the four matched-filter scalar
products $\{\vx\}_\mu \equiv \scalar{x}{h}$, similar to the $\F$-statistic \eqref{eq:31}.
The four $x_\mu$ are Gaussian distributed with mean and covariance obtained from \eqref{eq:5} as:
\begin{equation}
  \label{eq:62}
  \expect{\vx} = \vs\,,\qquad
  \cov{\vx}{\vx} = \M\,,
\end{equation}
We can decorrelate these variables by diagonalizing the detector response matrix $\M$ of
\eqref{eq:74} in the form
\begin{equation}
  \label{eq:107}
  \M = \R\,\W\,\R\transpose,
\end{equation}
in terms of the orthogonal rotation matrix
\begin{equation}
  \label{eq:73}
  \R = \left(\begin{array}{cc}
    r & 0\\
    0 & r\\
  \end{array}\right),
\quad\text{with}\quad
r \equiv \left(\begin{array}{cc}
  \cos\theta & -\sin\theta\\
  \sin\theta & \cos\theta\\
\end{array}\right),
\end{equation}
with the rotation angle $\theta$ given by
\begin{equation}
  \label{eq:108}
  \tan 2\theta = c,\quad\text{with}\quad  c\equiv\frac{2C}{A - B}\,.
\end{equation}
The resulting diagonal matrix is $\W=\dfact\,\diag{w^\plus,w^\cross,w^\plus,w^\cross}$ in terms of
the two unique eigenvalues
\begin{equation}
  \label{eq:105}
    w^\pluscross = \frac{1}{2}\left( A + B \pm (A-B)\sqrt{1+c^2}\right)\,.
\end{equation}
The action of $\R$ on the amplitude-vector $\Amp^\mu$ can be shown to be
\begin{equation}
  \label{eq:109}
  \vAmp' \equiv \R\transpose\vAmp = \left.\vAmp\right|_{2\psi\rightarrow 2\psi-\theta},
\end{equation}
namely a rotation of the polarization axes in the sky plane by $\theta$, defining a new polarization
angle $2\psi' \equiv 2\psi - \theta$, defining search-specific ``$\plus$'' and ``$\cross$'' polarizations.
It is further useful to define the matrix
\begin{equation}
  \label{eq:110}
  \Q \equiv \R\sqrt\W,
\end{equation}
such that
\begin{equation}
  \label{eq:114}
  \Q \Q\transpose = \M,
  \quad\text{and}\quad
  \Q\transpose\Q = \W\,.
\end{equation}
This allows us to define the transformed matched-filter quantities $z_\mu=\{\vz\}_\mu$ as
\begin{equation}
  \label{eq:111}
  \vz \equiv \Q\inv\vx,
\end{equation}
with mean and covariance
\begin{equation}
  \label{eq:112}
  \eqalign{
    \expect{\vz} &= \Q\inv\vs = \Q\transpose\vAmp = \sqrt\W\,\vAmp',\\
    \cov{\vz}{\vz\transpose} &= \Q\inv\M\Q\inv{}\transpose = \I\,,
  }
\end{equation}
i.e., the four $z_\mu$ are uncorrelated unit normal variates.

\subsection{Coherent statistics}
\label{sec:coherent-statistics}

The $\bcoh$-statistic of \eqref{eq:124} now reads as
\begin{equation}
  \label{eq:115}
  \eqalign{
    \bcoh(\vx) &\equiv \frac{\vx\transpose\vx}{\dfact} =
    \frac{\vz\transpose\W\vz}{\dfact}
     = w^\plus\,2\F_\plus(\vx)+ w^\cross\,2\F_\cross(\vx)\,,
  }
\end{equation}
where we defined
\begin{equation}
  \label{eq:116}
  2\F_\plus(\vx)\equiv z_1^2+z_3^2\,\qquad
  2\F_\cross(\vx)\equiv z_2^2 + z_4^2\,,
\end{equation}
which are two \emph{uncorrelated} $\chi^2$-distributed statistics with two degrees of freedom and
non-centrality parameters
\begin{equation}
  \label{eq:117}
  \eqalign{
    \rho^2_\plus  &\equiv \dfact w^\plus\left[\left(\Amp'^1\right)^2 + \left(\Amp'^3\right)^2\right]
    = \hrel^2\,w^\plus\,\alpha_1(\cosi,\psi'),\\
    \rho^2_\cross &\equiv \dfact w^\cross\left[\left(\Amp'^2\right)^2 + \left(\Amp'^4\right)^2\right]
    = \hrel^2\,w^\cross\,\alpha_2(\cosi,\psi'),
  }
\end{equation}
respectively, in terms of the intrinsic (rotated) polarization angle $\psi'$ introduced in the
previous section.
Thus $\bcoh(\vx)$ follows a \emph{generalized} $\gchi^2$-distribution
\cite{davies_distribution_1980,das_method_2021}, which we denote as
$\bcoh\sim\gchi_{\vec{w},\vec{2},\vec{\rho^2}}$, with vectors of weights
$\vec{w}=(w^\plus,w^\cross)$, degrees of freedom $\vec{2}=(2,2)$ and non-centrality parameters
$\vec{\rho^2}=(\rho^2_\plus,\rho^2_\cross)$.
This distribution has known mean and variance
\begin{equation}
  \label{eq:118}
  \eqalign{
    \expect{\bcoh} &= \sum_{p=+,\times}w^p \left(2+\rho^2_p\right),\\
    \var{\bcoh} &= 2\sum_{p=+,\times} (w^p)^2\left(2 + 2\rho^2_p\right)\,.
  }
\end{equation}

It is interesting to note that the $\F$-statistic \eqref{eq:31}, which in these variables reads as
\begin{equation}
  \label{eq:113}
  2\F(x) = \vx\transpose\M\inv\vx = \vz\transpose\vz = z_1^2+z_2^2+z_3^2+z_4^2\,,
\end{equation}
and therefore follows a $\chi^2$-distribution with four degrees of freedom as first shown in
\cite{jks98:_data}, is simply an \emph{unweighted} sum of the two partial statistics
$2\F_{\pluscross}$, namely
\begin{equation}
  \label{eq:120}
  2\F(x) = 2\F_\plus(\vx) + 2\F_\cross(\vx)\,,
\end{equation}
with corresponding mean and variance
\begin{equation}
  \label{eq:58}
  \eqalign{
    \expect{2\F} = 4 + \rho^2\,,\quad
    \var{2\F} = 2(4 + 2\rho^2)\,,
  }
\end{equation}
in terms of the total signal power
\begin{equation}
  \label{eq:121}
  \rho^2 = \rho_\plus^2 + \rho_\cross^2\,.
\end{equation}
This is effectively the unweighted special case $w^\plus=w^\cross=1$ of the $\bcoh$-statistic.

\subsection{Semicoherent $\bsco$-statistic}
\label{sec:semic-bsco-stat}

The semi-coherent generalization of the $\bcoh$-statistic \eqref{eq:125} can now be expressed as
\begin{equation}
  \label{eq:123}
  \bsco(x) \equiv \sum_{\ell=1}^{N\seg}w^\ell\bcoh_\ell(x) = \sum_{\ell=1}^{N\seg}\sum_{p=+,\times} w^{\ell{}p}\,2\F_{\ell{}p}(x),
\end{equation}
in terms of $2N\seg$ independent $\chi^2$-distributed statistics $2\F_{\ell{}p}$ with two degrees of
freedom and non-centrality parameters $\rho^2_{\ell{}p}$, respectively, and per-segment weights
$w^{\ell{}p}$ defined as
\begin{equation}
  \label{eq:6}
  w^{\ell{}p} \equiv w^\ell w^p_{(\ell)}\,,
\end{equation}
in terms of the polarization weights $w^p_{(\ell)}$ of \eqref{eq:105} for segment $\ell$ and the
data-weights $w^\ell$ of (\ref{eq:3}).
Therefore we see that $\bsco$ follows a generalized $\gchi^2$-distribution
$\gchi^2_{\vec{w},\vec{2},\vec{\rho^2}}$, with length-$2N\seg$ vectors
\begin{equation}
  \label{eq:94}
  \eqalign{
    \vec{w} &= (w^{1\plus}, w^{1\cross}, \ldots, w^{N\seg\plus}, w^{N\seg\cross}),\\
    \vec{2} &= (2,2, \ldots, 2,2),\\
    \vec{\rho^2} &= (\rho^2_{1\plus},\rho^2_{1\cross},\ldots,\rho^2_{N\seg\plus},\rho^2_{N\seg\cross}),
  }
\end{equation}
with resulting mean and variance
\begin{equation}
  \label{eq:38}
  \eqalign{
    \expect{\bsco} &= \sum_{\ell{}p}w^{\ell{}p}\left(2 + \rho^2_{\ell{}p}\right),\\
    \var{\bsco} &= 2 \sum_{\ell{}p} (w^{\ell{}p})^2\,\left(2 + 2\rho_{\ell{}p}^2\right).
  }
\end{equation}

\subsection{Noise distribution and false-alarm probability}
\label{sec:false-alarm-prob}

The mean and variance \eqref{eq:38} for $\bsco$ in the noise case can be made more explicit as
\begin{equation}
  \label{eq:39b}
  \eqalign{
    \left.\expect{\bsco}\right._{\hO=0} &= 2\sum_{\ell{}p} w^{\ell{}p} = \trace\sco{M},\\
    \left.\var{\bsco}\right._{\hO=0} &= 4 \sum_{\ell{}p} (w^{\ell{}p})^2
    = \frac{4}{\mean{\dfact}^2} \sum_{\ell} \dfact_\ell^2\left(A_\ell^2 + B_\ell^2 + 2 C_\ell^2\right),
  }
\end{equation}
where we used the eigenvalue expressions of \eqref{eq:105}.
Although the generalized $\gchi^2$-distribution has a known characteristic function, obtaining a
probability density function (pdf) or threshold-crossing probabilities generally requires either
direct numerical approaches (e.g., see \cite{davies_distribution_1980,das_method_2021}) or
Monte-Carlo simulation.

Some more analytical progress can be made in the noise case $\HypG: \hO=0$:
the per-segment quantities $Z_{\ell\plus}\equiv \sqrt{w^{\ell\plus}}(z_{\ell1} - i z_{\ell3})$ and
$Z_{\ell\cross}\equiv\sqrt{w^{\ell\cross}}(z_{\ell2}-i z_{\ell4})$ are then circularly-symmetric
(centered) complex Gaussians with zero mean, $\expect{Z_{\ell{}p}}=0$, and variances
$\expect{|Z_{\ell{}p}|^2}=2w^{\ell{}p}$, where $p\in\{+,\times\}$.
We can write the semi-coherent $\bsco$-statistic in these variables as
\begin{equation}
  \label{eq:127}
  \bsco(x) = \sum_{\ell{}p}|Z_{\ell{}p}|^2,
\end{equation}
and one can then show (cf.\ Eq.~(19) in \cite{hammarwall_acquiring_2008}) that the noise pdf is
\begin{equation}
  \label{eq:128}
  \prob{\bsco}{\HypG} = \sum_{\ell{}p} \frac{e^{-\frac{\bsco}{2w^{\ell{}p}}}}
  {2w^{\ell{}p} \prod_{\ell'p'\not=\ell{}p}\left(1-\frac{w^{\ell'p'}}{w^{\ell{}p}}\right)}\,,
\end{equation}
provided that all weights $w^{\ell{}p}$ are unique.
In practice this expression will be problematic for numerical reasons, as any weights being too
close to each other will lead to the denominator approaching zero.

Alternatively, in the limit of many segments, $N\seg\gg1$ one could use the central-limit Gaussian
distribution with given mean and variance \eqref{eq:38}, which will likely yield a feasible
approximation for many use cases, but testing and quantifying this approach left to future work.

In the coherent case we can write this as
\begin{equation}
  \label{eq:98}
  \prob{\bcoh}{\HypG} = \frac{1}{2(w^\plus - w^\cross)}\left[
    e^{-\frac{\bcoh}{2w^\plus}} - e^{-\frac{\bcoh}{2w^\cross}}
  \right],
\end{equation}
and analytically integrate for the false-alarm probability
\begin{equation}
  \label{eq:99}
  \eqalign{
    p\FA(\bcoh\thresh) &\equiv \prob{\bcoh>\bcoh\thresh}{\HypG}
    =\int_{\bcoh\thresh}^\infty \prob{\bcoh}{\HypG}\,d\bcoh\\
                   &=\frac{1}{w^\plus - w^\cross}\left[
                     w^\plus e^{-\frac{\bcoh\thresh}{2w^\plus}}
                     - w^\cross e^{-\frac{\bcoh\thresh}{2w^\cross}}
                     \right],
  }
\end{equation}
in terms of the false-alarm threshold $\bcoh\thresh$.
This result agrees with the analysis in \cite{astone_method_2014} (see Eqs.(34) and (35)) for the
classic 5-vector statistic, which is equivalent to $\bcoh$-statistic as mentioned in
Sec.~\ref{sec:weak-signal-bayes} and shown explicitly in \ref{sec:relation-5-vector}.

Contrary to the $\chi^2$-distributed statistics such as $2\F$ or the dominant-response statistics
$\F\sub{AB}$, the noise distribution (and therefore false-alarm probabilities) of the
$\bcoh$-statistic depends on the sky position due to the antenna-pattern weights $w^\pluscross$.
The same is true for the weighted $\F\sub{w}$ and $\F\sub{ABw}$ statistics of \eqref{eq:43}.
This sky-dependent false-alarm probability complicates the application of this statistic to all-sky
searches and we postpone this topic to future work, focusing instead on searches in single
phase-evolution points $\dop$ instead.  We have verified the qualitative robustness of all
subsequent results by testing in different sky points (not shown).

The agreement between the theoretical noise distribution \eqref{eq:98} and the measured distribution
for $\bcoh$ will be tested in the next section.

\section{Tests and numerical results}
\label{sec:numerical-results}

\subsection{Synthesizing statistics and $\chi^2$ sensitivity estimates}
\label{sec:test-method}

Most of the following tests use \emph{synthesized} statistics, a method first introduced in
\cite{2009CQGra..26t4013P,2014CQGra..31f5002W}, which consists of drawing samples for the 4-vectors
$x_\mu$ from its 4-dimensional Gaussian distribution with mean
$s_\mu=\scalar{s}{h_\mu} = \M_{\mu\nu}\Amp^\nu$ and covariance $\M_{\mu\nu}$ for signal amplitude
parameters $\Amp^\mu$.
Given all the statistics tested here are functions of these four $x_\mu$, this allows us to
efficiently generate samples for the corresponding statistics without requiring a full dataset $x$
or computing explicit matches $x_\mu=\scalar{x}{h_\mu}$.

For the $\F$- and dominant-response $\F\sub{AB}$ statistics, however, an even more efficient (and
accurate) method consists in directly computing false-alarm and detection probabilities by
numerically integrating the $\chi^2$-distributions governing these statistics, see
\cite{wette2011:_sens,2018arXiv180802459D}.
When using this $\chi^2$-based sensitivity estimation method, we denote the corresponding statistics
as $\F\chisq$ and $\F\sub{AB}\chisq$ in the legend, to distinguish them from the (default)
synthesized sampling method.

For the following results with synthesized statistics, we use $N\sub{noise}=\num{e7}$ noise samples
(to estimate false-alarm thresholds), and $N\sub{signal}=\num{e6}$ signal+noise samples for
estimating the detection probabilities.
When using the $\chi^2$-integration method instead, we use $\num{e5}$ samples for $\rho^2$ at fixed
$\hO$, histogrammed into $\num{1000}$ bins for the numerical sensitivity integrals.

While all following plots of detection probability show $90\%$ uncertainty bands, these tend to be
smaller than the line-width and are therefore generally not visible.

\subsection{Noise distribution of coherent $\bcoh$-statistic}
\label{sec:noise-distr-coher}

\newcommand{\metaNumTrials}{1000000}
\newcommand{\metaTStart}{\SI{1234567890}{s}}
\newcommand{\metaTSpan}{\SI{900}{\second}}
\newcommand{\metaDetectors}{H1,L1}
\newcommand{\metaNumSegments}{1}
\newcommand{\metaSkyA}{(5.16,0.78)\,\si{\radian}}
\newcommand{\metaWeightA}{(0.02,0.23)}
\newcommand{\metaSkyB}{(0.32,0.49)\,\si{\radian}}
\newcommand{\metaWeightB}{(0.01,0.82)}

\begin{figure}[htbp]
  \centering
  \includegraphics[width=0.8\textwidth]{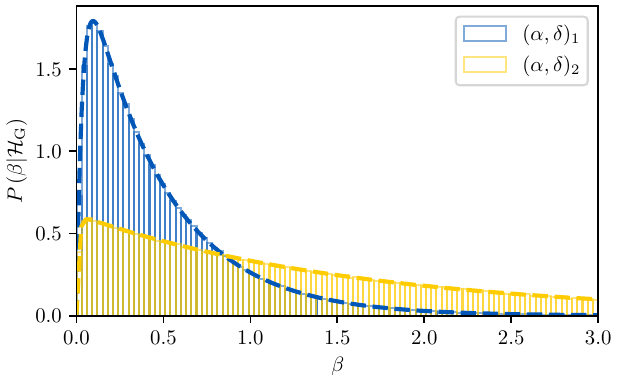}
  \caption{Noise distribution of coherent $\bcoh$ for $T\seg=\metaTSpan$ (starting at
    GPS time $t_0=\metaTStart$, data from \metaDetectors{}), for two sky positions
    $(\alpha,\delta)_1=\metaSkyA$, and $(\alpha,\delta)_2=\metaSkyB$.
    The corresponding weights are
    $(w^\plus,w^\cross)_1=\metaWeightA$ and $(w^\plus,w^\cross)_2=\metaWeightB$, respectively.
    Dashed lines show the theoretical generalized $\gchi^2$-distribution of \eqref{eq:98}, and the
    histograms are computed on $\num[scientific-notation]{\metaNumTrials}$ synthesized values for
    $\bcoh$.
  }
  \label{fig:noise-dist}
\end{figure}

Figure \ref{fig:noise-dist} shows the noise-distribution of the
coherent $\bcoh$-statistic for two different sky positions $(\alpha,\delta)_{12}$ on a short
segment of data $T\seg=\metaTSpan$ from two detectors (H1 and L1).
We see excellent agreement between the histogrammed synthesized $\bcoh$-values and the theoretical
generalized-$\gchi^2$ noise-distribution of \eqref{eq:98}.

\subsection{Coherent statistics}
\label{sec:coherent-statistics-1}

\begin{figure}[htbp]
  \centering
  case (i) H1, $T\seg=\SI{25}{h}$, $(\alpha,\delta)=(2,-0.5)\,\si{rad}$\\
  \includegraphics[width=0.8\textwidth]{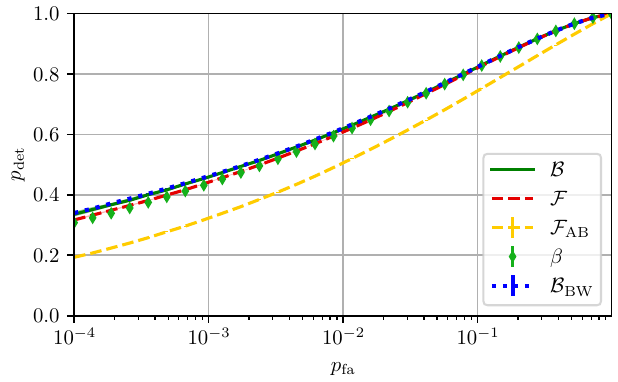}\\
  case (ii) H1, $T\seg=\SI{86164}{\second}$, $(\alpha,\delta) = (2, 0)\,\si{rad}$\\
  \includegraphics[width=0.8\textwidth]{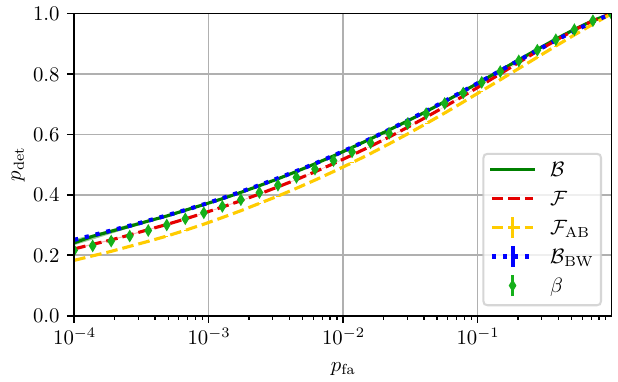}\\
  case (iii) H1+L1, $T\seg=\SI{100}{\second}$, $(\alpha,\delta)=(2,-0.5)\,\si{rad}$\\
  \includegraphics[width=0.8\textwidth]{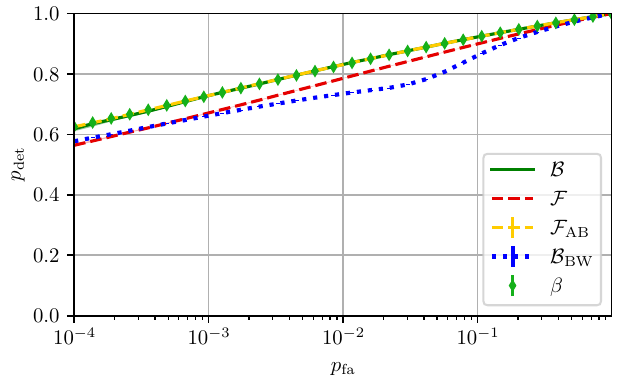}\\
  \caption{
    Detection probability as a function of false-alarm probability for the three different coherent
    test cases (i)-(iii) previously considered in \cite{2019CQGra..36a5013B}.
    In all cases the signal population has a fixed relative amplitude of $\hrel=10$.
  }
  \label{fig:bero-whelan-i-iii}
\end{figure}

In Fig.~\ref{fig:bero-whelan-i-iii} we present the same three coherent test cases used in
\cite{2019CQGra..36a5013B}, extended with the dominant-response $\F\sub{AB}$-statistic \eqref{eq:60}
of \cite{covas_improved_2022}, and the weak-signal Bayes-factor approximation $\bcoh$ of
\eqref{eq:124}.
All three cases use signals at fixed (relative) amplitude of $\hrel=10$.
\begin{enumerate}
\item[(i)] \emph{top panel}: a coherent $T\seg=\SI{25}{h}$ search on H1 data (starting at GPS time
  $\SI{756950413}{\second}$) for sky-position $(\alpha,\delta)=(2,-0.5)\,\si{rad}$ (the original
  example from \cite{2009CQGra..26t4013P}).
  Here $\bcoh$ performs similarly to $\F$ and slightly worse than $\B\BeroWhelan$, which perfectly
  approximates the performance of $\B$.
  The dominant-response $\F\sub{AB}$ performs worst as expected in this relatively long-segment case.

\item[(ii)] \emph{middle panel}: a similar case with
  $T\seg=\SI{1}{\textrm{sidereal day}}$ ($\SI{86164}{\second}$), same start-time as (i), for a
  single detector (H1) and sky-position at the equator $(\alpha,\delta)=(2,0)\,\si{rad}$, with
  qualitatively similar results case~(i).

\item[(iii)] \emph{bottom panel}: a short coherence length of $T\seg=\SI{100}{\second}$ for two
  detectors (H1 and L1), sky position $(\alpha,\delta)=(2,-0.5)\,\si{rad}$ at sidereal time $00:00$
  (using GPS mid-time $\SI{756581823}{\second}$).
  Here $\B\BeroWhelan$ performs worse or similar to $\F$ (reproducing the result of
  \cite{2019CQGra..36a5013B}), while both the dominant-response $\F\sub{AB}$ as well as the
  weak-signal $\bcoh$-statistics perform as well as the full $\B$-statistic
\end{enumerate}

Note that these results serve mostly for comparison and validation purposes against previous results
on $\B$ and $\B\BeroWhelan$ shown in \cite{2019CQGra..36a5013B} and \cite{2009CQGra..26t4013P}.
Coherent searches of such short segment lengths (even of a day) are not very interesting from a
practical point of view.
Fully coherent searches would typically be over a much longer segments, or would be part of a
semi-coherent search.
For a more systematic and realistic study on the dependence on $T\seg$ we therefore move directly
to the semi-coherent cases shown in the next section.

% Dependence on $T\seg$:
% \begin{figure}[htbp]
%   \centering
%   \includegraphics[width=0.8\textwidth]{plotPdetVsTseg-rocs__plotPdetVsTseg__setup_coherent_True_DataCatalog_detectors_H1.pdf}
%   \includegraphics[width=0.8\textwidth]{plotPdetVsTseg-rocs__plotPdetVsTseg__setup_coherent_True_DataCatalog_detectors_H1_L1.pdf}
%   \includegraphics[width=0.8\textwidth]{plotPdetVsTseg-rocs__plotPdetVsTseg__setup_coherent_True_DataCatalog_detectors_H1_L1_V1.pdf}
%   \caption{Coherent detection probability (at fixed false-alarm
%     $p\FA=\num{e-3}$) as a function of segment length $T\seg$.}
%   \label{fig:pdet-vs-tseg}
% \end{figure}

\subsection{Semi-coherent results}
\label{sec:semi-coher-results}

We consider semi-coherent searches with a fixed total duration of $\Tspan=\SI{10}{\day}$, with
variable segment lengths spanning from a very short-segments at $T\seg=\SI{900}{\second}$ (with
$N\seg=960$ segments) to a fully-coherent search at $\Tspan=\SI{10}{\day}$.
Here we ensure that all $N\seg$ segments are of equal length $T\seg$, in order to avoid confounding
the results with the effect of varying data-factors shown and discussed in the next section.

\begin{figure}[htbp]
  \centering
  H1\\
  \includegraphics[width=0.8\textwidth]{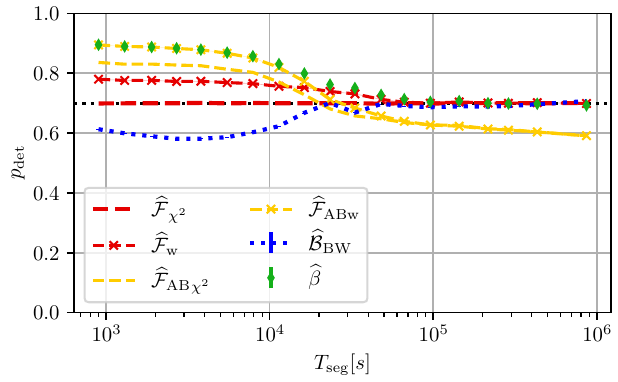}\\
  H1+L1\\
  \includegraphics[width=0.8\textwidth]{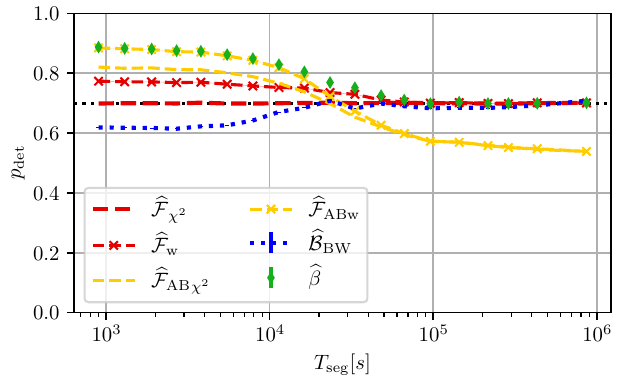}\\
  H1+L1+V1\\
  \includegraphics[width=0.8\textwidth]{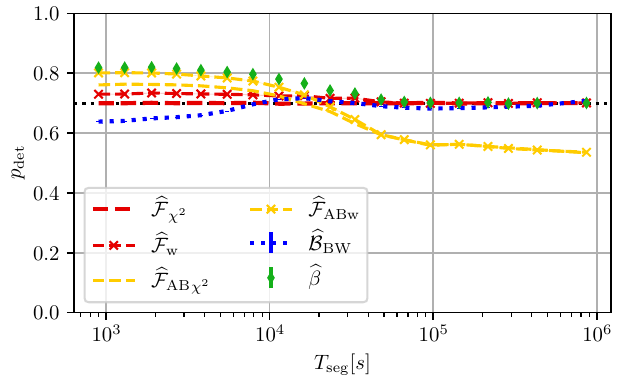}
  \caption{Detection probability as a function of semi-coherent segment length $T\seg$ at fixed
    total span of $\Tspan=\SI{10}{\day}$ (with GPS start time $\SI{756950413}{\second}$),
    false-alarm probability $p\FA=\num{e-3}$ and sky position $(\alpha,\delta)=(2,-0.5)\,\si{rad}$.
    The top panel is for a single detector (H1), middle panel is H1+L1 and bottom panel includes
    Virgo H1+L1+V1. The relative signal amplitude $\hrel(T\seg)$ is adjusted for a fixed detection
    probability $p\DET(2\sco{\F})=0.7$ for the standard semi-coherent $\sco{\F}$-statistic.  }
  \label{fig:res-semi-coherent}
\end{figure}

The results of these simulations are shown in figure~\ref{fig:res-semi-coherent}.
We use a fixed false-alarm probability of $p\FA=\num{e-3}$ and a sky position of
$(\alpha,\delta)=(2,-0.5)\,\si{rad}$.
Note that here we do not include a semi-coherent summed $\B$-statistic, as this requires numerical
2D integration for each sample and segment, making it impractical to obtain good sampling in a
reasonable time. This is of course the same reason this statistic is not practical for any real
searches in the first place.

The signal strength $\hrel(T\seg)$ is determined for each $T\seg$ by requiring a fixed detection
probability of $p\DET(2\sco{\F})=0.7$ for the standard semi-coherent $\sco{\F}$-statistic, in order
to operate in a relevant $p\DET$ range across all $T\seg$ despite the greatly different
sensitivities.
This results in a range of $\hrel(T\seg=\SI{900}{\second})\sim1.9$ to
$\hrel(T\seg=\SI{10}{\day})\sim13.9$ for the \emph{per-segment} relative amplitude.
As discussed in Sec.~\ref{sec:do-detect-sign}, even at the shortest segment length this
relative signal amplitude does \emph{not} fall into the ``weak-signal'' regime of $\hrel\ll1$.

The top panel in figure~\ref{fig:res-semi-coherent} shows the single-detector case for H1, the
middle panel is for H1+L1 and bottom panel additionally includes Virgo V1, with qualitatively
similar results in all three cases:
\begin{itemize}
\item[-] the $\B\BeroWhelan$ approximation performs well at segment lengths $T\seg\gtrsim\SI{1}{\day}$,
  where it effectively equals the performance of the (weighted and unweighted) $\sco{\F}$
  statistics.
  This may seem surprising given Fig.~\ref{fig:bero-whelan-i-iii}, but at the larger $p\DET$ and
  $p\FA$ values used here the differences become very small.
  As anticipated from the coherent results and the discussion in \cite{2019CQGra..36a5013B},
  $\B\BeroWhelan$ increasingly loses sensitivity at shorter segment lengths (and can run into
  similar numerical degeneracy issues with inverting $\M_{\mu\nu}$ \eqref{eq:25} as the
  $\F$-statistic \cite{covas_improved_2022}).

\item[-] the dominant-response statistic $\sco{\F}\sub{AB}$ performs better than $\sco{\F}$ at short
  $T\seg$ and increasingly poorly at longer segments $T\seg\gtrsim\SI{19}{\hour}$, as expected
  \cite{covas_improved_2022}.
  We can also confirm that the (empirical) weighting scheme \eqref{eq:43} introduced in
  \cite{covas_improved_allsky_2022} further improves the short-segment performance, for \emph{both}
  $\sco{\F}\sub{w}$ as well as $\sco{\F}\sub{ABw}$.
  This can be understood from the fact that the antenna-pattern sensitivity to a fixed sky position
  varies more strongly over segments when they are short, which is where the segment weighting can
  gain more sensitivity.
  When the antenna-pattern response is averaged over longer segments $\gtrsim\Ord{\si{\day}}$, it
  will be increasingly constant over segments, which is why segment weighting has less of an
  effect there.

  However, this Monte-Carlo simulation uses perfectly white Gaussian noise of constant
  noise floor and $\SI{100}{\percent}$ duty factor over all segments, which is why the contribution
  of the data factor $\dfact_\ell$ to the segment weights does not confer any benefits at longer
  segment durations $T\seg\gtrsim\SI{1}{\day}$ here, which will be discussed further in the next
  section.

\item[-] the semi-coherent weak-signal Bayes-factor approximation $\bsco$ performs effectively
  ``optimally'' across both the short- and long-segment regimes, equalizing (and improving upon)
  the weighted dominant-response statistic  $\sco{\F}\sub{ABw}$ at short segments, while converging
  to the $\sco{\F}\sub{(w)}$ and $\B\BeroWhelan$ performance at long segments (at least for
  equal-data-factor segments, see next section).
\end{itemize}

\subsection{Performance for unequal data-factors $\dfact_\ell$}
\label{sec:perf-uneq-data}

The per-segment weights of various semi-coherent weighted statistics $\sco{\F}\sub{w}$,
$\sco{\F}\sub{ABw}$, as well as the weak-signal Bayes-factor approximation $\bsco$ are a combination
of an antenna-pattern- and a data-factor weight contribution, see section~\ref{sec:semi-coher-gener}.

The semi-coherent Monte-Carlo simulations in the previous section use ideal segments with equal
noise-floors and data amounts, and therefore equal per-segment data factors $\dfact_\ell$.  Given
that for segments longer than a day the antenna-pattern weight contributions also become
increasingly constant, this explains why the weighted statistics $\sco{\F}\sub{w}$,
$\sco{\F}\sub{ABw}$ and $\bsco$ do not confer any benefits over the unweighted version in that case,
as seen in figure~\ref{fig:res-semi-coherent}.

\begin{figure}[htbp]
  \centering
  \includegraphics[width=0.8\textwidth]{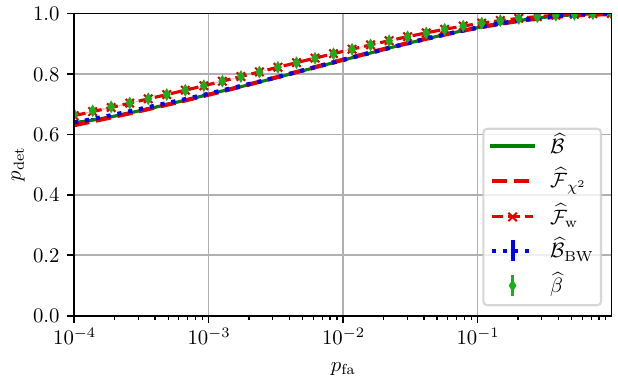}
  \caption{Detection probability as a function of false-alarm probability for a semi-coherent search
    on three unequal-$\dfact_\ell$ segments of $T\seg=\SI{25}{h}$, with per-segment duty factors of
    $0.1$, $1$ and $1$, respectively, assuming two detectors H1 and L1.  The signal sky-position is
    $(\alpha,\delta)=(2,-0.5)$ and the relative amplitude is fixed at $\hrel=10$.  }
  \label{fig:roc-unequal-dfact}
\end{figure}

As a proof of principle we now consider a rather extreme example of a semi-coherent search with
three segments of $T\seg=\SI{25}{\hour}$ span with widely different duty factors of $[0.1, 1, 1]$,
respectively.
In other words, the first segment has data gaps and only contains $\SI{10}{\percent}$ of the
data compared to the other two segments with no gaps.  This results in the same relative variation
of per-segment data factors $\dfact_\ell$.
The result is shown in figure~\ref{fig:roc-unequal-dfact}, illustrating that the weighted statistics
$\sco{\F}\sub{w}$ and $\bsco$ now do outperform both $\sco{\F}$ and $\B\BeroWhelan$, as well as
(maybe somewhat surprisingly) the full $\B$-statistic.

This result seems to strongly suggest that it is specifically the \emph{weak-signal} (half-Gaussian)
prior of \eqref{eq:190} (but also \eqref{eq:8}) that yields the right per-segment weighting, while
the strong-signal limit $\B$-statistic (and its approximation $\B\BeroWhelan$) intrinsically lack this feature.

\section{Conclusions}
\label{sec:conclusions}

In this paper we have generalized the Bayesian $\B$-statistic by using a half-Gaussian
$\hO$-prior instead of a uniform prior.
This reduces to the $\B$-statistic in the strong-signal limit ($H\rightarrow\infty$) but yields a new analytic
approximation in the weak-signal limit ($H\rightarrow0$), defining the new weak-signal statistic
$\bcoh(x)$.
This statistic is shown to follow a generalized $\gchi^2$-distribution with $(2,2)$ degrees of
freedom per segment, with weights determined by the eigenvalues of the detector-response matrix
$\M_{\mu\nu}$ and non-centrality parameters given by the corresponding signal-power components.

The sensitivity of $\bcoh$ is found to be comparable to $\F$ for day-long coherent searches, making
it slightly less sensitive than the $\B$-statistic (and the Bero-Whelan approximation).
However, for semi-coherent searches, it matches or outperforms both
(i) the best short-segment statistic (i.e., the weighted dominant-response statistic
$\sco{\F}\sub{ABw}$), and
(ii) the best practical long-segment statistic, i.e., the weighted $\sco{\F}\sub{w}$-statistic.

Overall, the weak-signal $\bcoh$-statistic appears to be a powerful and practical detection
statistic across a wide range of segment lengths that we have tested, namely from
$\SI{900}{\second}$ to $\SI{10}{days}$.
Further testing is needed across various real-world scenarios and different false-alarm
regimes beyond those considered here.

One interesting practical question is how to correctly perform an all-sky search with a statistic
(such as $\bcoh$) that has a sky-position-dependent false-alarm distribution.
This issue is not new and already affects any antenna-pattern-weighted segment statistics.
A common practical solution is to use a ``noise-normalized'' critical-ratio statistic for the
candidate ranking.

Another potentially interesting application of this new statistic is the line-robust framework of
\cite{keitel_search_2014,keitel_robust_2016}.
The prior scaling parameter $H$, which governs the new Bayes factor, has a more direct physical
interpretation compared to the unphysical prior cutoff parameter in the $\F$-statistic-prior, which
complicates the correct tuning of the line-robust $B\sub{S/GL}$ statistics.

\ack
I am very grateful to Pep B.~Covas for numerous insightful discussions on short-segment statistics
that ultimately motivated this work.
I thank Luca D'Onofrio for very helpful feedback on the 5-vector framework translation.

\appendix

\section{Alternative derivation of $\bcoh$ from weak-signal $\F$-statistic prior}
\label{sec:altern-deriv-weak}

Similar to the idea of using a (half-) Gaussian $\hO$-prior \eqref{eq:190},
we can use a Gaussian $\Amp^\mu$ prior centered on zero with scale $H$, namely
\begin{equation}
  \label{eq:8}
  \prob{\Amp}{\HypS} = \frac{1}{(2\pi)^2 H^4} e^{-\frac{1}{2}\Amp^\mu C_{\mu\nu}\Amp^\nu}\quad
  \text{with}\quad
  C_{\mu\nu} \equiv \frac{1}{H^2}\delta_{\mu\nu}\,,
\end{equation}
which reduces to the original uniform-$\Amp^\mu$ $\F$-statistic prior of Sec.~\ref{sec:recov-f-stat}
in the strong-signal limit $H\rightarrow\infty$.
This prior preserves the functional form of the likelihood ratio \eqref{eq:35} and we can write the
resulting Bayes-factor integral in analogy to \eqref{eq:44} as
\begin{equation}
  \label{eq:9}
  B\sub{\Signal/\Gauss}(x) = \frac{1}{\sqrt{\det\M\sub{H}}} e^{\F\sub{H}(x)}\,,
\end{equation}
with modified detector-response matrix
\begin{equation}
  \label{eq:10}
  \M\sub{H}{}_{\mu\nu} \equiv \M_{\mu\nu} + \frac{1}{H^2}\delta_{\mu\nu}
  = \dfact \left(\begin{array}{cc}
    m\sub{H} & 0\\
    0 & m\sub{H}
    \end{array}\right),
\end{equation}
where
\begin{equation}
  \label{eq:12}
  m\sub{H} \equiv \left(\begin{array}{cc}
    A + \frac{1}{\dfact H^2} & C\\
    C & B + \frac{1}{\dfact H^2}
  \end{array}\right)
\end{equation}
and we defined the modified $\F\sub{H}$-statistic as
\begin{equation}
  \label{eq:13}
  2\F\sub{H}(x)\equiv x_\mu \left(\M\sub{H}\inv\right)^{\mu\nu} x_\nu\,.
\end{equation}

Considering $\det\M\sub{H} = \dfact^4 D\sub{H}^2$ with the sub-determinant
\begin{equation}
  \label{eq:14}
  D\sub{H} \equiv \det m\sub{H} = D + \frac{A + B}{\dfact H^2} + \frac{1}{\dfact^2H^4}\,,
\end{equation}
in the weak-signal limit $\Hrel^2\ll 1$ we obtain the limiting expressions
$D\sub{H}\overset{{}H\rightarrow0}{\longrightarrow} \frac{1}{\dfact^2{}H^4}$,
$\M\sub{H}\rightarrow \frac{1}{{}H^2}\I$, and $\sqrt{\det\M\sub{H}}\rightarrow H^{-4}$, and finally
\begin{equation}
  \label{eq:16}
  2\F\sub{H}(x) \overset{{}H\rightarrow0}{\longrightarrow} \dfact{}H^2 \bcoh(x),
\end{equation}
which yields the resulting Bayes factor as
\begin{equation}
  \label{eq:34}
  B\sub{S/G}\rightarrow H^4 e^{\frac{1}{2}\Hrel^2\,\bcoh(x)}\,,
\end{equation}
in terms of the weak-signal $\bcoh$-statistic of \eqref{eq:124}.
While normalized differently, this is otherwise equivalent to the weak-signal Bayes factor obtained
in \eqref{eq:1}.

\section{General matched-filter scalar product $\scalar{x}{y}$}
\label{sec:weight-multi-detect}

The scalar product $\scalar{x}{y}$ of \eqref{eq:29} can be fully generalized
\cite{prix:_cfsv2,2014CQGra..31f5002W} to several detectors $X=1,\ldots, \Ndet$, each providing a
finite set of data chunks (commonly referred to as short Fourier transforms or SFTs)
$\alpha=1,\ldots, N\sft^X$, each of length $\Tsft$, allowing for gaps in between SFTs and for
individual noise-floors $\Sn_{X\alpha}$ (assumed stationarity only during each SFT).
This results in the general form for the scalar product
\begin{equation}
  \label{eq:24}
  \scalar{x}{y} = 2 \dfact\,\av{x\,y},\quad\text{with}\quad\dfact \equiv \Sn\inv\, \Tdata,
\end{equation}
where the dimensionless \emph{data factor} \cite{covas_improved_2022} $\dfact$ is defined in terms
of the \emph{overall} noise-floor $\Sn$ and total amount of data $\Tdata$, namely
\begin{equation}
  \label{eq:32}
  \Sn\inv \equiv \frac{1}{N\sft}\sum_{X\alpha} \Sn_{X\alpha}\inv\,,\qquad
  \Tdata \equiv N\sft\,T\sft\,,
\end{equation}
and $N\sft\equiv \sum_X N\sft^X$. The weighted multi-detector average can be written as
\begin{equation}
  \label{eq:28}
  \av{x\,y} \equiv \frac{1}{N\sft}\sum_{X\alpha} w_{X\alpha}\,\frac{1}{\Tsft}\int_0^{\Tsft} x_{X\alpha}(t)\, y_{X\alpha}(t)\,dt,
\end{equation}
where $x_{X\alpha}$ refers to the data chunk $\alpha$ from detector $X$, and the corresponding noise
weights are defined as $w_{X\alpha}\equiv \Sn\inv_{X\alpha}/\Sn\inv$, such that $\sum_{X\alpha} w_{X\alpha}=N\sft$.

\section{Matched filter in physical coordinates}
\label{sec:scal-prod-phys}

Expressing the scalar-product ``match term'' $\scalar{x}{h} = \Amp^\mu x_\mu$ of \eqref{eq:18} in
physical amplitude coordinates $\{\hO,\cosi,\psi,\phiO\}$ yields (after some tedious but
straightforward algebra):
\begin{equation}
  \label{eq:173}
  \eqalign{
    \scalar{x}{h} &= \Amp^\mu x_\mu = Q\sub{s}\,\sin\phiO + Q\sub{c}\,\cos\phiO\\
                 &= Q(x; \hO,\cosi,\psi) \, \cos\left(\phiO - \varphi_0\right)\,,
                 }
\end{equation}
with
\begin{equation}
  \label{eq:174}
  \eqalign{
    Q\sub{s} &\equiv -\sin2\psi\left(x_1\,\Across + x_4\,\Aplus\right)
    + \cos2\psi\left(x_2\,\Across - x_3\,\Aplus \right)\,,\\
    Q\sub{c} &\equiv \hspace{1em}\sin2\psi\left(x_2\,\Aplus - x_3\,\Across\right)
    + \cos2\psi\left(x_1\,\Aplus + x_4\,\Across\right)\,,
    }
\end{equation}
where $\Aplus \equiv \hO\,( 1 + \cosi^2)/2$ and $\Across \equiv \hO\,\cosi$, and where we defined
\begin{equation}
  \label{eq:17}
  \eqalign{
    Q^2 &\equiv Q\sub{s}^2 + Q\sub{c}^2 = \hrel^2q^2\,,\\
    \tan\varphi_0 &\equiv {Q\sub{s}}/{Q\sub{c}},
    }
\end{equation}
which was used to write the likelihood ratio in the form \eqref{eq:42}.

\section{Relation to 5-vector-method statistic}
\label{sec:relation-5-vector}

Here we discuss the detailed translation of the formalism and relation of the $\bcoh$-statistic
of \eqref{eq:124} to the original 5-vector statistic of \cite{astone_method_2010}, which was also
discussed recently in \cite{donofrio_two_2024}.
Starting with Eqs.~(2.1) and (2.2) in \cite{2014CQGra..31f5002W} for the gravitational-wave tensor
$\tens{h}(t)$, which we can write as
\begin{equation}
  \label{eq:7}
  \tens{h}(\tau) = \Re\left\{ \left(\Aplus\,\tens{e}_\plus -i \Across\,\tens{e}_\cross\right)e^{i\Phi(\tau)}\right\},
\end{equation}
in terms of the signal phase $\Phi(\tau)=\phi(\tau)+\phiO$ and the intrinsic wave basis tensors $\tens{e}_{\plus\cross}$.
Using the relations Eqs.~(A.3),(A.4) in \cite{astone_method_2010}, namely
$\five{h}_0=\sqrt{\Aplus^2+\Across^2}$ and $\five{\eta}\equiv-\Across/\Aplus$\footnote{Not to be
  confused with our definition $\cosi\equiv\cos\iota$. For clarity we decorate all
  5-vector-specific quantities with $\five{\;}$ in case there is ambiguity.}, we obtain
\begin{equation}
  \label{eq:37}
  A_\plus = \frac{\five{h}_0}{\sqrt{1+\five{\eta}^2}}\,,\quad
  A_\cross = \frac{-\five{\eta}\,\five{h}_0}{\sqrt{1+\five{\eta}^2}}\,,
\end{equation}
and using the $\psi$-independent set of wave basis tensors $\mathfrak{e}_{\plus\cross}$ via
Eq.~(2.5) in \cite{2014CQGra..31f5002W}, we can rewrite \eqref{eq:7} as
\begin{equation}
  \label{eq:48}
  \tens{h}(\tau) = \Re\left\{\five{h}_0\left(\five{H}_\plus\,\mathfrak{e}_\plus +
        \five{H}_\cross\,\mathfrak{e}_\cross \right) e^{i\Phi(\tau)}\right\},
\end{equation}
yielding Eq.~(1) in \cite{astone_method_2010} and Eqs.~(2),(3) defining the complex 5-vector
amplitudes $\five{H}_{\plus\cross}$.

The strain $h(t)$ measured in the detector is obtained via contraction (in both indices) with the
detector tensor $\tens{d}$ as $h(t) = \tens{h}:\tens{d}$, resulting in Eq.~(9) in
\cite{astone_method_2010}, namely
\begin{equation}
  \label{eq:50}
  h(t) = \Re\left\{\five{h}_0\left(\five{A}^\plus(t)\,\five{H}_\plus +
        \five{A}^\cross(t)\,\five{H}_\cross\right) e^{i\Phi(t)}  \right\},
\end{equation}
with
\begin{equation}
  \label{eq:51}
  \five{A}^\plus(t) \equiv \mathfrak{e}_\plus : \tens{d} \equiv a(t)\,,\quad
  \five{A}^\cross(t)\equiv \mathfrak{e}_\cross : \tens{d} \equiv b(t)\,,
\end{equation}
establishing an important equivalence with the ``JKS'' antenna-pattern functions $a(t),\,b(t)$
defined in Eq.~(2.12) in \cite{2014CQGra..31f5002W} as used here, e.g., \eqref{eq:basisfunc}.

A key feature in the 5-vector formalism is to explicitly write the time dependence of the
antenna-pattern functions $\five{A}^{\plus\cross}(t)$ in terms of the five harmonics of the sidereal
angular rotation rate $\Omega$ of the Earth, namely\footnote{Simplified by dropping a constant phase
  factor $e^{ik(\alpha-\beta)}$ (for detector longitude $\beta$) that can be absorbed into a shift
  in $t$.}.
\begin{equation}
  \label{eq:52}
  \five{A}^{\plus\cross}(t) = \vec{A}^{\plus\cross}\cdot e^{-i\vec{k}\Omega\,t}\,,
\end{equation}
with $\vec{\;\;}$ denoting 5-component vectors, and $\vec{k}\equiv(-2,-1,0,1,2)$.
The data five-vector $\vec{X}$ defined in Eq.~(20) of \cite{astone_method_2010} is
\begin{equation}
  \label{eq:54}
  \vec{X} \equiv \int_T x(t)\, e^{-i\vec{k}\Omega t} e^{-i\phi(t)}dt\,,
\end{equation}
and using the above translations, the two scalar products $\vec{X}\cdot\vec{A}^{\plus\cross}$ can
be expressed in our notation as
\begin{equation}
  \label{eq:55}
  \eqalign{
    \vec{X}\cdot\vec{A}^\plus &= \int_T x(t) a(t) e^{-i\phi(t)}dt = \frac{\Sn}{2}\left(x_1 - ix_3\right)\,,\\
    \vec{X}\cdot\vec{A}^\cross &= \int_T x(t) b(t) e^{-i\phi(t)}dt = \frac{\Sn}{2}\left(x_2 - ix_4\right)\,,\\
  }
\end{equation}
in terms of the noise PSD $\Sn$ and the data scalar products $x_\mu$ defined in \eqref{eq:41},
and where we used the fact that $\vec{A}^{\plus,\cross}\cdot e^{-i\vec{k}\Omega t}=a(t),b(t)$, see Eqs.~\eqref{eq:51},\eqref{eq:52}.
We note that these two terms correspond exactly to JKS $F_{ab}$ defined in \cite{jks98:_data} Eqs.(97),(98).
The original 5-vector statistic $\five{S}_0$ introduced in \cite{astone_method_2010} Eq.(25) (with
coefficients $\five{c}_{\plus\cross}=|\vec{A}^{\plus\cross}|^4$) is now expressible as
\begin{eqnarray}
  \label{eq:56}
  \five{S}_0 &\equiv |\vec{X}\cdot\vec{A}^\plus|^2 + |\vec{X}\cdot\vec{A}^\cross|^2\nonumber\\
             &= \frac{\Sn^2}{4}\left( x_1^2+x_2^2+x_3^2+x_4^2\right)\nonumber\\
  &=\frac{\Sn T}{4}\,\bcoh(x)\,,
\end{eqnarray}
in terms of the weak-signal statistic $\bcoh(x)$ defined in \eqref{eq:124}.

Considering the translation of the ``JKS'' antenna-pattern matrix coefficients A,B,C of
\eqref{eq:74}, using Eqs.~\eqref{eq:51} and \eqref{eq:52} we find
\begin{equation}
  \label{eq:59}
  \eqalign{
    A\equiv\av{a^2} &= \sum_{kl=-2}^{2}\vec{A}^{\plus}_k\vec{A}^{\plus*}_l\,\mathcal{E}_{(k-l)}(T),\\
    B\equiv\av{b^2} &= \sum_{kl}\vec{A}^{\cross}_k\vec{A}^{\cross*}_l\,\mathcal{E}_{(k-l)}(T),\\
    C\equiv\av{a\,b} &= \sum_{kl}\vec{A}^{\plus}_k\vec{A}^{\cross*}_l\,\mathcal{E}_{(k-l)}(T),\\
  }
\end{equation}
in terms of the (complex) sinc function 
\begin{equation}
  \label{eq:61}
  \mathcal{E}_{\Delta k}(T) \equiv \frac{1}{i\Delta k \Omega T}\left(e^{i\Delta k\Omega T} - 1\right)
  = e^{i\pi\Delta k \,T/T\sid}\,\mathrm{sinc}\left(\Delta k\, \frac{T}{T\sid}\right)\,.
\end{equation}
In two special cases, namely (i) integer multiple sidereal days
$T=n\,T\sid$, or (ii) long durations $T\gg T\sid$, we find
$\mathcal{E}_{\Delta k}\rightarrow
\delta_{\Delta k}$ in terms of the Kronecker delta, and therefore
\begin{equation}
  \label{eq:64}
  (i),(ii): \quad A\rightarrow|\vec{A}^\plus|^2,\;
  B\rightarrow |\vec{A}^\cross|^2,\;
  C\rightarrow \vec{A}^\plus\cdot\vec{A}^{\cross*} = 0\,.
\end{equation}
Therefore, considering the second 5-vector statistic $\five{S}_1$ defined in
\cite{astone_method_2010}, namely
\begin{eqnarray}
  \label{eq:69}
  \five{S}_1&\equiv \frac{|\vec{X}\cdot\vec{A}^\plus|^2}{|\vec{A}^\plus|^2} + \frac{|\vec{X}\cdot\vec{A}^\cross|^2}{|\vec{A}^\cross|^2}\nonumber\\
  &= \frac{\Sn^2}{4}\left(\frac{x_1^2+x_3^2}{|\vec{A}^\plus|^2} +
    \frac{x_2^2+x_4^2}{|\vec{A}^\cross|^2}\right),
\end{eqnarray}
we find in the the equivalence (in special cases (i),(ii)):
\begin{equation}
  \label{eq:70}
  \five{S}_1 \stackrel{(i),(ii)}{\longrightarrow}\frac{\Sn T}{4}\left[2(\F\sub{A} +
    \F\sub{B})\right] = \frac{\Sn T}{4} 2\F(x)\,,
\end{equation}
using definitions \eqref{eq:40} and \eqref{eq:71} with the fact that
$D\equiv AB-C^2 \stackrel{(i),(ii)}=AB $.  This agrees with the statement in
\cite{astone_method_2010} and \cite{donofrio_two_2024} that this statistic is the $\F$-statistic,
although we are able to show this only in the special cases (i) and (ii) above.

\section*{References}

\bibliography{bib}

\end{document}